\documentclass[twocolumn]{aastex63}

\shorttitle{Li Evolution in NGC 6819}
\shortauthors{Deliyannis, Anthony-Twarog, Lee-Brown, Twarog}

\begin{document}


\title{Li Evolution and the Open Cluster NGC 6819: A Correlation Between Li Depletion and Spindown in Dwarfs More Massive Than the F-Dwarf Li-Dip}


\author{Constantine P. Deliyannis}
\affiliation{Department of Astronomy, Indiana University, Bloomington, IN 47405-7105, USA }
\email{cdeliyan@indiana.edu}

\author{Barbara J. Anthony-Twarog}
\affiliation{Department of Physics and Astronomy, University of Kansas, Lawrence, KS 66045-7582, USA}
\email{bjat@ku.edu}

\author{Donald B. Lee-Brown}
\altaffiliation{Insight Data Science, New York, NY 10016}
\affiliation{Department of Physics and Astronomy, University of Kansas, Lawrence, KS 66045-7582, USA}
\email{dbleebrown@gmail.com}

\author{Bruce A. Twarog}
\affiliation{Department of Physics and Astronomy, University of Kansas, Lawrence, KS 66045-7582, USA}
\email{btwarog@ku.edu}



\begin{abstract}
Spectroscopy of 333 NGC 6819 stars and {\it Gaia} astrometry are used to map Li evolution from the giant branch tip to 0.5 mag below the Li dip. Isochrone comparison with $[Fe/H] = -0.04$, based upon neural network spectroscopic analysis, produces an age of 2.25 (2.4) Gyr for $E(B-V)$ = 0.16 (0.14) and $(m-M)$ = 12.40 (12.29). Despite originating outside the Li dip, only 10\% of single subgiants/giants have measurable Li. Above the Li dip, the limiting A(Li) for single stars is 3.2 $\pm$ 0.1 but the lower range is comparable to that found within the dip. The F-dwarf Li dip profile agrees with the Hyades/Praesepe, evolved forward. The Li level among stars populating the plateau fainter than the Li dip is A(Li) = 2.83 $\pm$ 0.16; the dispersion is larger than expected from spectroscopic error alone. Comparison of Li and $V_{ROT}$ distributions among turnoff stars in NGC 7789, NGC 2506, NGC 3680, and NGC 6819 indicates that rotational spindown from the main sequence is critical in defining the boundaries of the Li dip. For higher mass dwarfs, spindown is likewise correlated with Li depletion, creating a second dip, but at higher mass and on a longer time scale. The Li distribution among evolved stars of NGC 6819 is more representative of the older M67, where subgiant and giant stars emerge from within the Li dip, than the younger NGC 7789, where a broad range in $V_{ROT}$ among the turnoff stars likely produces a range in mass among the giants.
\end{abstract}



\section{Introduction}
The challenges to understanding stellar evolution are numerous. For the majority of stars change occurs at a prohibitively slow pace, necessitating comparisons among supposedly similar stars of different ages, under circumstances where the degree of similarity may be questionable and the relative ages indeterminate. Direct observation of stars is limited to the surface properties, requiring inference based upon theoretical models of the stellar interior to confirm or contradict the plausibility of the models. While asteroseismology has provided an increasingly reliable probe of the structure and evolutionary state of stars within the {\it Kepler} field \citep[see, e.g.][]{ST11, WU14, WL14, SI15, HA17}, the more traditional approach to assess what goes on beneath the surface has relied on detecting the changes wrought by the mixing of partially processed materials from the interior to the stellar atmosphere. Among the elements adopted for this purpose, Li has proven invaluable due its propensity for destruction above a well-defined temperature and the expectation that it will vary over time for any star where a convective or mixed envelope of sufficient depth can access the Li-depleted stellar interior. This surface signature can be altered by extending the mixing zone deeper, by extending the Li-depleted zone higher, or by the right combination of both.


Because they have the potential to minimize the range of variables that need to be considered, as well as supplying more precise values for those that do, star clusters remain an ideal environment for testing all aspects of stellar evolution, not just Li variation. 
However, when it comes to Li evolution, comparison between theory and observation has not been kind to the models.  It is strikingly clear that at least some of the physical mechanisms left out of the Standard Stellar Evolution Theory (SSET), such as rotation, diffusion, mass loss or gain, magnetic fields, etc., significantly affect the surface Li abundance. 
Understanding the nature of the shortcomings of the SSET can help inform us about which additional mechanisms are important.  Perhaps the best success of the SSET comes from the agreement between the predicted degree of subgiant Li dilution in metal-poor stars \citep{DDK} and observations of field \citep{RY95} and cluster \citep{LI09} subgiants.  For main sequence stars, observation supports only qualitatively the SSET prediction \citep{DDK, PI97} that lower-mass stars have deeper surface convention zones (SCZs) and have thus depleted more Li.  Quantitatively, the best case scenario for the SSET is the {\it possible} agreement between the models and slowly-rotating cluster dwarfs with ages less than about 150 Myr (e.g. see discussion of the Pleiades in \citet{AT18b, CU17, SOP15}).  However, whereas the SSET predicts that stars that are now G dwarfs depleted their Li only during the early pre-main-sequence, real open cluster G dwarfs continue to deplete their Li during the main sequence (MS) \citep{JE97, SR05, CU17}.
On average, the older the star the worse the discrepancy, with the Sun being one of the most egregious offenders, having depleted a factor of 50 more Li than the factor of $\sim 3$ predicted by the SSET \citep{KI97, PI97, AS09,TH17}.
Another example is that of lower mass stars in the pre-main sequence and early MS phase, which exhibit large dispersions in Li at the same mass and age, with rapid rotators exhibiting higher Li abundances \citep{SOP14, SOP15, BO18, AT18b}.
A particularly distressing example is F dwarfs, in which deep envelope SCZs should not exist.  In sharp contradistinction to the SSET, F dwarfs develop severe Li depletions during the MS near $T_{\mathrm{eff}} =$ 6600 K, a phenomenon commonly known as the Li Dip \citep{BT86}.

Making sense of these discrepancies occupies an important role in mapping out post-MS evolution, and much progress has been made. 
Rotation-induced radius inflation is the leading contender to explain the large Li dispersions in young G and K dwarfs \citep{SOP15,AT18b, JDJ18}. For the non-SSET Li depletion in F and G dwarfs, a variety of evidence points to rotationally-induced mixing due to instabilities triggered by angular momentum loss \citep{PI90} as the dominant mechanism \citep{CU17}. This evidence includes the Li/Be depletion ratio \citep{DBS98,BAK04} the Be/B depletion ratio \citep{BO05, BO16}, higher Li in Short-Period-Tidally-Locked-Binaries 
(SPTLB) \citep{RY95, CU17}, the timing of the Li depletion \citep{SD04}, and the steepening of the Li-rotation relation with age in F dwarfs \citep{ST03}, among others; diffusion and slow mixing due to gravity waves might also play a role.

Understanding these discrepancies may also play a crucial role in cosmology.  The Spite Li plateau \citep{SP82a, SP82b} among the older, more metal-poor dwarfs populating the Galactic halo highlights a discrepancy of about a factor of three between the Li abundances of these stars and the inferred Big Bang Li value from Planck \citep{CO14}, {\it if} we assume that these stars have not depleted their Li. However, until we have a better handle on diffusion, mixing, and Li-destroying processes among lower mass stars of all metallicities, such claims seem premature \citep[e.g.][]{NO12, GR13, GR14, GR16}.
For example, although direct evidence remains elusive, rotationally-induced mixing is a very reasonable way to deplete Li in these stars by a factor of three.


With the goal of using atmospheric Li to probe stellar structure and evolution among low mass stars, the authors have undertaken an extensive spectroscopic program to survey members of a key set of open clusters from the tip of the giant branch to as far down the main sequence as the technology allows. Results have been published for the clusters NGC 3680 (age = 1.75 Gyr) \citep{AT09}, NGC 6253 (3.0 Gyr) \citep{AT10, CU12}, and, most recently, the metal-deficient open cluster NGC 2506 (1.85 Gyr) \citep{AT16, AT18a}. The current investigation reports on the analysis of over 330 stars in the older open cluster, NGC 6819 (2.3 Gyr). The cluster is defined by a unique combination of characteristics. Its location within the {\it Kepler} field has made it the focus of asteroseismic studies reaching down the giant branch \citep{ST11}, with a rapidly expanding literature related to the cluster and its members \citep{AT13, JE13, PL13, YA13, WU14, WL14, MI14, LB15, BR16, HA17}. The age of NGC 6819 situates it in a key evolutionary phase where the red giants come from stars on the hotter, supposedly undepleted, side of the Li dip, but the turnoff stars are still in a mass range where partial degeneracy at hydrogen exhaustion slows the evolutionary rate enough to populate both the subgiant branch and the first-ascent giant branch below the red giant clump, a trait it shares with the slightly younger but metal-deficient NGC 2506 \citep{AT18a}. Preliminary spectroscopic analysis of the sample discussed in this investigation led to the discovery of a unique Li-rich giant fainter than the level of the clump \citep{AT13}, below the point where standard stellar evolution models predict the initiation of mixing assumed to create Li-rich atmospheres \citep{CH10}. Since Li-rich stars at any location along the giant branch remain rare, at present their existence requires either a relatively specialized and restricted mixing or mass loss process or the merger of a planet of significant mass with its companion star \citep{CA16, AG16}. Spectroscopic \citep{CA15} and asteroseismic \citep{HA17} evidence suggests that this Li-rich star has a substantially lower mass than other cluster members in close propinquity on the HR diagram, which might suggest a severe He-core-flash at the RGB tip as the origin of both the extra Li and the mass loss. As we will discuss below, uncertainty about its cluster membership has been eliminated by the astrometric information supplied by the {\it Gaia} DR2 release \citep{GA18a}.

In addition to the Li-rich giant, astrometric \citep[][hereinafter PL]{PL13}, photometric \citep[][hereinafter Paper I]{AT14}, and spectroscopic \citep[][hereinafter Paper II]{LB15} investigations have found the cluster to have slightly subsolar metallicity, confirmed below from a new analysis using a neural network approach and by the high-dispersion spectroscopic work of \citet{SL19}, in contrast with claims of [Fe/H] comparable to the Hyades from earlier analysis of three red giants \citep{BR01}, and to be affected by variable reddening. The latter discovery is relevant because traditional high dispersion spectroscopic analysis requires reliable input parameters for the models used in interpreting the spectra. Stellar temperatures, if derived from photometric indices, and surface gravities, if derived using precise estimates of the cluster distance via comparison of the observed color-magnitude-diagram (CMD) to theoretical isochrones of appropriate age and metallicity, are both dependent upon the assumed reddening. Fortunately, a neural network approach has the capability of circumventing these issues.

The outline of the paper is as follows: Section 2 summarizes the spectroscopic data, discussed in detail in Paper II, and revisits the cluster membership taking the {\it Gaia} DR2 astrometry for NGC 6819 into account.  Section 3 lays out the reddening corrections, age and distance estimates through comparison of the CMD to theoretical isochrones. Section 4 reanalyzes the cluster metallicity using a neural network approach to the spectroscopy, and details the parameters leading to the spectroscopic Li abundances. Section 5 explores the patterns among the NGC 6819 Li abundances for the dwarfs and giants, while Section 6 discusses the trends among Li and the rotational distributions of various clusters. 
Section 7 summarizes our conclusions.

\section{Spectroscopic Observations and Data Reduction}
\subsection{Observations} 
NGC 6819 was an open cluster targeted for comprehensive analysis, including identification of cluster members, by the WIYN Open Cluster Study \citep{MA00}. Our initial spectroscopic sample of probable cluster members was constructed using the radial-velocity survey of NGC 6819 by \citet[][hereinafter H09]{HO09}. All stars brighter than $V \sim 16.75$ with radial-velocity membership probabilities greater than 50\% were identified as spectroscopic candidates, while stars classed as double-lined spectroscopic binaries were eliminated. Single-lined systems were retained since the existence of the companion would have minimal impact on spectral line measurement. Stars were not eliminated based upon their position in the CMD to avoid biasing the sample against stars undergoing potentially anomalous evolution.     

Spectroscopic data were obtained using the WIYN 3.5-meter telescope\footnote{The WIYN Observatory was a joint facility of the University of Wisconsin-Madison, Indiana University, Yale University, and the National Optical Astronomy Observatory.} and the Hydra multi-object spectrograph over 13 nights from September and October 2010, June 2011 and February 2013. Six configurations were designed to position fibers on a total of 333 stars. Detailed discussion of the processing and reduction of these spectra is presented in Paper II.
\subsection{Cluster Membership - Radial Velocities}
Comparison of our radial velocities ($V_{RAD}$) for 304 likely single-star members with those of H09 showed excellent agreement, with a difference of -0.27 km s$^{-1}$, in the sense $(H09 - Paper II)$, and a dispersion consistent with the predicted scatter from the individual measurements (Paper II). \citet{MI14} updated their high-precision radial-velocity studies in NGC 6819, so we have revised the comparison of our radial velocities (Paper II) to the expanded data set with essentially the same result. From over 300 single stars common to the two surveys, our radial velocities are larger by $0.2 \pm 1.1$ km s$^{-1}$ (sd), confirming the minor offset discussed in Paper II, but an insignificant difference with respect to either the variance among the residual values or the estimated error for a single star's radial velocity in our study, 1.1 km s$^{-1}$. As noted in Paper II, the comparison using single-lined binaries shows a dramatically larger offset and scatter, as expected.

The discussion by \citet{MI14} incorporates membership probabilities using both the proper motions of PL, who provided membership information for over 15,000 stars in this rich cluster field with the highest precision within 10\arcmin\ of the cluster center, and the distribution of the WOCS radial velocities. Since our sample was compiled prior to PL, it relied heavily upon the radial-velocity work of H09, with the result that of the 333 stars in our sample, only 1 has a radial-velocity membership probability below 50\%. Not surprisingly, the astrometric work of PL tagged 59 of the remaining 332 stars as proper-motion nonmembers, eliminating these interlopers from the spectroscopic abundance analysis of Paper II. With the advent of {\it Gaia} \citep{GA16}, a new level of precision has been added to the astrometric database, requiring a reevaluation of the earlier, ground-based astrometric classifications.
\subsection{Cluster Membership - {\it Gaia}}
To identify potential cluster members within NGC 6819, we will follow the simple prescription adopted in \citet{AT18b}. While {\it Gaia} DR2 astrometry has been used by \citet{CA18} to select highly probable astrometric members, only a handful of these stars have {\it Gaia} DR2 radial velocities, and a subset of our stars have poor to inadequate {\it Gaia} DR2 astrometric measures. Stars will be classified initially as proper-motion members or nonmembers based upon their position within the proper-motion vector-point diagram, taking into account the dispersion among the cluster members and the individual uncertainty in the measured proper motions. From the identified members, a second check is made using the derived cluster parallax, eliminating those stars which deviate from the cluster mean value by more than three times the quoted uncertainty in the parallax. As seen below, this simple approach is more than adequate for our current needs.

Cross-matching our spectroscopic sample with the {\it Gaia} DR2 catalog, all stars were found but four retained only coordinate positions and no astrometric information (5006, 10010, 14002, 16005; all numbers refer to the ID on the WOCS system, HO09). For these stars, we have defaulted to the membership classification from the radial-velocity and proper-motion probabilities compiled by \citet{MI14}; all are probable members. 

As a first cut on the {\it Gaia} DR2 sample, the quoted uncertainties in the positions for each of the 329 stars were combined to identify stars where the astrometry was likely to be unreliable since the positional errors invariably translated into large uncertainties in the proper motion and parallax. Nine stars (4008, 7004, 11014, 13007, 15002, 22005, 22020, 47007, 49023) were found to have combined positional errors above 0.1 mas. From PL, 7004, 15002 and 22005 are nonmembers and will be eliminated.  Of the 6 remaining members (PL), three (4008, 13007, 22020) are SB1 binaries \citep{MI14}.

To define the cluster reference motion and parallax, the {\it Gaia} DR2 cross-match was restricted to 
190 stars with both high-precision DR2 proper motions and PL probabilities above 90\%. 
Mean cluster values of the proper motion in both right ascension and declination were determined and the radial vector distance of each star from the cluster mean calculated. Eight stars with total proper motion placing them more than 0.30 mas yr$^{-1}$ away from the cluster proper-motion vector point were removed and the centroid rederived. The adopted cluster proper-motion center is -2.9159 $\pm$ 0.0072 (sem) mas yr$^{-1}$ and -3.8584 $\pm$ 0.0074 (sem) mas yr$^{-1}$ in right ascension and declination, respectively. If we then restrict our parallax sample to only 123 stars within 0.15 mas yr$^{-1}$ of the cluster proper motion, the mean cluster parallax becomes 0.3552 $\pm$ 0.0025 (sem) mas. As expected, these are all in excellent agreement with the values derived by \citet{CA18} from 1589 probable (above 50\%) members from astrometry alone.  
These will be adopted for the cluster in the discussion which follows. 
We note that there is significant evidence that the {\it Gaia} DR2 parallaxes suffer from a zero-point error, leading to an underestimate of the parallax and overestimate of the distance, a point we will return to in detail in Section 3.

Returning again to the 320 stars with reliable {\it Gaia} DR2 coordinates, 26 stars with proper-motion vectors placing them more than 0.6 mas yr$^{-1}$ away from the cluster motion were classed as nonmembers. Of the 26, 19 have $\Delta$$\pi$/$\sigma$$_\pi$ $>$ 3, where $\Delta$$\pi$ is defined as the absolute value of ($\pi_{clus}$ - $\pi_{star}$), confirming in part the field-star classification. Of the 26, only 2 had PL proper-motion membership above 31\% and and 20 had 0\% probability.

Among the proper-motion distribution, 17 stars have proper-motion vectors which place them radially between 0.3 and 0.6 mas yr$^{-1}$ away from the cluster motion. Only 2 of these stars have $\Delta$$\pi$/$\sigma$$_\pi$ $>$ 3. Of the 17, PL assigns proper-motion membership at 98\% or above for 12. The remaining 5 are at 14\% or less, including both stars which are probable parallax nonmembers; only the the parallax nonmembers will be excluded from our final sample.

Of the remaining 277 stars with proper-motion vectors within a radial vector distance of 0.3 mas yr$^{-1}$ from the cluster's motion, only 7 have  $\Delta$$\pi$/$\sigma$$_\pi$ $>$ 3; these will be excluded from the membership list. This last set of 270 members includes 25 stars with ground-based proper-motion probabilities  below 50\% and 206 with probabilities at or above 90\% (PL); 20 stars are SB1 binaries. The full set of probable members sits at 295, of which 26 are SB1. Table 1 contains a complete listing of the membership classification for each star based upon the {\it Gaia} DR2 data, detailing if it is consistent with membership via proper motion, parallax, or both. Only stars meeting both criteria will be treated as probable cluster members in the discussions which follow.
\subsection{WOCS 7017 - Li-Rich Giant}
\citet{CA15} and \citet{HA17} have discussed the evidence for and against the membership of the Li-rich giant, 7017, in NGC 6819 with admirable thoroughness. As suggested by \citet{AT13}, the large errors associated with PL's proper motion for this fascinating star implied that the older, more positive estimation of membership probability provided by \citet{S72} should keep 7017 on the table as a possible cluster member. \citet{CA15} strengthened that case by providing an explanation for the atypically large error in PL's proper motion, as well as bolstering the membership credentials of 7017 through an independent spectroscopic estimation of the star's gravity and abundance pattern. An intriguing feature of their discussion of this Li-rich giant was the derivation of a spectroscopically-based, anomalously low mass, possibly related to its surface Li abundance, but confirming the structural uniqueness of this red clump giant demonstrated by asteroseismology \citep{ST11, HA17}. The analysis above places 7017 in the category of definite member from both proper motion and parallax.

\section{Stellar Properties: Variable Reddening, Isochrone Ages, and Distance Moduli}
One important challenge for any investigation of the stellar properties in NGC 6819 is the variable reddening across the face of the cluster, as demonstrated by PL. The sense and amplitude of the variable reddening were validated in Paper I; Fig. 10 of that paper showed the tightening of the Str\"omgren CMD that results from applying broadly-derived spatial reddening estimates to colors and magnitudes, while Fig. 11 used broadband photometry \citep{RV98} and Yale-Yonsei \citep[$Y^2$;][]{DE04} isochrones for the determination of age and distance. For Figs. 10 and 11 of Paper I, individual deviations from the average foreground reddening were determined for three spatial zones, with appropriate adjustments to the photometric indices, e.g. individual $(B-V)$ colors were corrected by an amount equal to $\delta$$E_{(B-V)}$, defined as $E(B-V)_{star} - 0.16$, with a corresponding adjustment of $3.1 \delta$$E_{(B-V)}$ to the $V$ magnitudes, placing all stars under a uniform reddening of $E(B-V)$ = 0.16, the cluster mean as derived from extended Str\"omgren photometry of stars at the turnoff (Paper I). For future reference, $(B-V)'$ will denote $(B-V)-$$\delta$$E_{(B-V)}$ and $V'$ will represent $V - 3.1$$\delta$$E_{(B-V)}$. The $V$ and $(B-V)$ data for the spectroscopic sample are the same as compiled and discussed in Paper II. For the additional members used to define the CMD, especially below the magnitude limit of the spectroscopy, $V$ magnitudes are taken from Paper I, while the initial $(B-V)$ indices are those of \citet{RV98}, the same system used as the standard for the merger of the multiple color sources in Paper II.
\subsection{Isochronal Constraints}
Fig. 1 of Paper II emphasized the CMD locations for the spectroscopic sample, with different symbols indicating membership and binarity from H09. Fig. 1 of the current paper provides a dramatic update, making use of the radial-velocity membership determinations from \citet{MI14} and the astrometric analysis of Section 2. Large symbols designate the stars of Table 1 with blue open symbols, black crosses, red solid squares, and black solid squares denoting probable single-star members, nonmembers, member binaries, and nonmember binaries, respectively. For each star in the spectroscopic sample, individual reddening estimates were derived from the reddening map of PL, with values listed in Table 1 of Paper II. 

To extend the sample beyond the depth limits of our spectroscopy while minimizing the impact of the variable reddening, stars within 6\arcmin\ of the cluster center were processed through the same {\it Gaia} DR2 astrometric procedure as the spectroscopic sample. Because of increasing astrometric errors at fainter magnitudes, there is an artificial decline in the number of stars retained as members, with a cutoff near $V$ $\sim$ 17.5. Those stars bright enough to be included in \citet{MI14} were checked and all stars classed as nonmembers or uncertain members were eliminated. All stars were individually reddening corrected and plotted as small open triangles in Fig. 1; no distinction is made between single stars or binaries. We emphasize that the membership selection is optimized to delineate the cluster CMD sequence, so eliminating nonmembers is a clear  priority over completeness. 

\begin{figure}
\figurenum{1}
\plotone{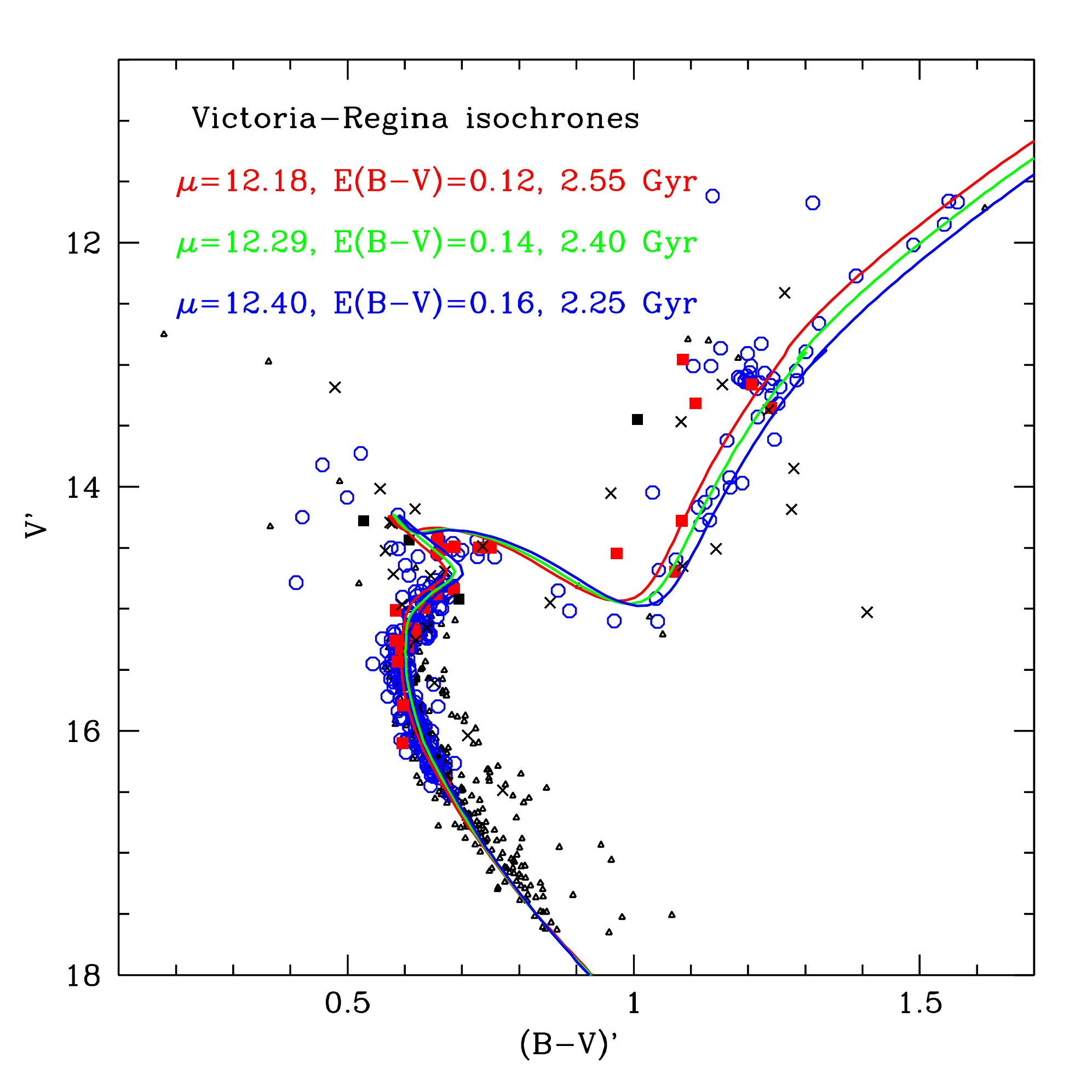}
\caption{Color-magnitude diagram of NGC 6819 with the effects of variable reddening removed. Larger symbols designate stars from Table 1. Solid squares are single-lined spectroscopic binaries, red for members and black for field stars. Blue open circles are single, probable members and black crosses denote single, probable nonmembers. Small triangles include only astrometric members from {\it Gaia} DR2, with only definite radial-velocity nonmembers removed. Isochrones from the VR compilation are shown, with $[Fe/H]$ = -0.04 and three assumed reddening values, $E(B-V)$ = 0.12 (red), 0.14 (green) and 0.16 (blue), with respective ages and apparent moduli of 2.55 Gyr and 12.18, 2.4 Gyr and 12.29, and 2.25 Gyr and 12.40.}
\end{figure}

For the comparison of Fig. 1, we have adopted the isochrones from \citet[][hereinafter VR]{VR06}, constructed for an abundance of $[Fe/H] = -0.04$, rather than the $Y^2$ models of \citet{DE04} from Papers I and II. The adopted $[Fe/H]$ is the quadratic sum of results from Papers I and II, but has been confirmed through the use of a neural network abundance analysis applied to the high-resolution spectra which form the basis of the current investigation, as discussed in Section 4. Because of its location within the {\it Kepler} field, the NGC 6819 CMD, age, and distance modulus have garnered significantly more discussion than most clusters. While a comprehensive overview of the cluster properties is neither necessary nor desirable for our purposes, a few points regarding the CMD match should be made.	

While use has been made of the {\it Gaia} DR2 parallaxes for individual stars in weighing the probability of cluster membership, the success of this technique is almost wholly dependent upon the relative precision of the astrometric measures and tells us little, if anything, about potential systematics in the parallax scale. If $(m-M)$ = 12.40 and $A_V$ = 0.5, then $(m-M)_o$ = 11.90, $d$ = 2400 pc, and $\pi$ = 0.417 mas, noticeably larger than obtained above (0.355 mas) using a simple average of highly probable members. A more elaborate approach as illustrated by \citet{CA18} derives the identical result within the errors, $\pi$ = 0.356 mas. However, as discussed by \citet{ST18, ZI18} and \citet{RI18}, there is growing evidence for a zero-point offset to the {\it Gaia} DR2 parallax scale at the level of 0.05 to 0.08 mas, with the offset size potentially dependent upon position in the sky. \citet{CA18}, through comparison with cluster distances derived by the BOCCE project \citep{BR06}, find a typical systematic offset to the {\it Gaia} DR2 parallaxes equal to -0.05 mas. Applying this to the data for NGC 6819 produces $\pi$ = 0.405 mas, in excellent agreement with the main sequence fit given the uncertainty in the parallax zero-point.

Keeping in mind that the detailed intermediate-band analysis of the cluster (Paper I) defines $E(B-V)$ = 0.16 as the cluster mean reddening, we have compiled two alternative matches defined by  lower reddening to illustrate the trend. The three isochrones shown in Fig. 1 have been adjusted for an optimum fit to the turnoff and unevolved main sequence under the assumption that the reddening is $E(B-V)$ = 0.16 (blue curve), 0.14 (green curve), or 0.12 (red curve). We have incremented the VR $(B-V)$ colors by +0.01, in conformity with our past usage of isochrones zeroed to a solar color of $(B-V)$ = 0.65 at an age of 4.6 Gyrs \citep[e.g.][]{TW09, RA12b}. The apparent moduli have been set to assure identical fits to the observed main sequence at $(B-V)' = 0.85$. As expected, the highest reddening leads to the youngest age (2.25 Gyr vs. 2.55 Gyr) and a larger apparent distance modulus (12.40 vs. 12.18). While the subgiants appear too faint relative to the models, independent of the adopted reddening, the isochrones of higher reddening nicely bracket the giant branch from the base to above the level of the clump. We note that the fit of the VR isochrone to the photometry is essentially identical to that presented in Paper I, in which $Y^2$ isochrones were used; with $E(B-V)$ = 0.16 and $[Fe/H] = -0.06$, an age of 2.3 to 2.5 Gyr was derived for an apparent modulus of 12.4. 

Use of a higher adopted metallicity for the cluster would require an even younger isochrone and a larger distance modulus. This prediction can be tested using the analysis of multiple isochrone sets and NGC 6819 in Fig. 10 of \citet{JE13}, who assumed $E(B-V)$ = 0.12, $(m-M)$ = 12.3 and $[Fe/H]$ = +0.09; if an isochrone of the exact metallicity was not available, the one closest to +0.09 was selected. For VR, the closest match was $[Fe/H]$ = +0.13. With their lower reddening (0.12) relative to 0.16, our apparent modulus from Fig. 1 is 12.18; partial compensation comes from a metallicity higher by 0.17 dex, which should boost the modulus by $\sim$0.17 mag \citep{TW09}, leading to a final value of $(m-M)$ = 12.35, the same within the uncertainties as adopted by \citet{JE13}. In agreement with our Fig. 1, the color of their turnoff best matches their age of 2.25 Gyr at the lower reddening due to the higher metallicity, while the subgiants appear fainter than their models at this age. Note also that their unevolved main sequence lies increasingly above the cluster photometry as one moves down the main sequence, while the isochrone fit in Fig. 1 remains consistently within or at the lower edge of the distribution at fainter magnitudes. This difference reflects the changing slope of the main sequence with changing metallicity.

By contrast, the $Y^2$ match \citep{JE13} indicates an age midway between 2.25 and 2.5 Gyr, with a good fit from the lower main sequence through to the subgiant branch. The result should be the same for higher reddening and lower metallicity, with the expectation that the unevolved main sequence models should lie increasingly fainter than the photometry at fainter magnitudes, as confirmed in Fig. 11 of Paper I. For completeness, the BaSTI isochrones \citep{PI04} with $[Fe/H] = +0.06$ supply a good match from the unevolved main sequence through the subgiant branch for an age of 2.25 Gyr, while the DSEP isochrones \citep{DO08} imply an age between 2.75 and 3.0 Gyr from the turnoff, with the subgiant models too faint compared to the photometry.  For isochrone comparisons based upon $VI$ rather than $BV$, the reader is referred to Fig. 8 of \citet{BR16}. 

As is obvious, age and distance estimates through isochrone fitting depend strongly upon the adopted reddening and metallicity, as well as the choice of isochrones. For fixed metallicity and reddening, the latter becomes the dominant source of scatter among investigations. As noted earlier, we have consistently attempted to minimize the impact of different approaches to stellar models, to different transformations of the models from the theoretical to the observational plane, and to differences in the assumed solar composition by requiring that a one-solar-mass star with $[Fe/H] = 0.0$ have a specific $B-V$ and $M_{V}$ at an age of 4.6 Gyr. Such simple zero-point offsets become less reliable as $[Fe/H]$ deviates from solar but, as exemplified by the comparison of the results from the $Y^2$ and VR isochrones for NGC 6819, they can lead to greater consistency in both age and distance. 
\subsection{Eclipsing Binary Constraints}
An alternative which minimizes the role of metallicity and the adopted isochrones is the use of eclipsing binaries, as detailed in the exquisite analysis of three systems in NGC 6819 by \citet{BR16}, an expansion and revision of the earlier work by \citet{JE13} and \citet{SA13}. With radius and $T_{\mathrm{eff}}$ known from the binary analysis and photometric temperatures, one can derive $M_{V}$ with minimal impact due to reddening and/or metallicity uncertainty. The weighted average in $(m-M)$ is 12.38 $\pm$ 0.04, where the quoted uncertainty is the error in the mean, in excellent agreement with the original estimate from Paper I and the VR fit in Fig. 1 for $E(B-V)$ = 0.16.  

Equally important, with the masses and radii known, \citet{BR16} derive an age for NGC 6819 through an extensive set of isochrone and model comparisons. To account for the possibility that the metallicity could be either approximately solar or higher by $\sim$0.1 dex, the analysis is done under two different assumptions for the composition. A key difference in their approach is the method by which they determine the choice of isochrones. To avoid concerns about the assumed Z$_{\sun}$ for each isochrone source, the isochrones selected from each set are picked to have Z = 0.012 and 0.015, irrespective of the adopted Z$_{\sun}$ for the models. The isochrones are normalized, independent of distance and reddening, by forcing a match between the observed position of 24009 C in the CMD and a star of the same mass at a given age as defined by the isochrones. In effect, 24009 C takes the equivalent role of the sun in fixing the position of the isochrones to the CMD. The age of the cluster can then be set by mapping how well the other members of the binary systems and the evolved stars at the top of the turnoff match the predicted position of the isochrones. Using this CMD-based approach, \citet{BR16} derive an age of 2.21 $\pm$ 0.10 $\pm$ 0.20 Gyr for the cluster, in excellent agreement with the fit in Fig. 1. The agreement is relevant because the adopted isochrone match will be used to define the relationship between the stars' positions within the CMD and the predicted masses used to delineate the trends of Li with age and mass, as discussed below.

However, the challenges posed by the differences in the construction of theoretical isochrones and their transformation to the observational plane can be seen in the range of values obtained by using the forced match of the isochrones to the CMD to derive $E(B-V)$ and $(m-M)$. For Z = 0.015, \citet{BR16} find $E(B-V)$ between 0.19 and 0.22 and $(m-M)$ between 12.46 and 12.57; for Z = 0.012, both $E(B-V)$ and $(m-M)$ are systematically larger by 0.03 and 0.1, respectively.

Returning to the distribution of stars on the CMD, the inclusion of the astrometric constraints imposed by {\it Gaia} DR2 has significantly reduced the scatter in the CMD, both among the spectroscopic sample and the fainter main sequence. Of the spectroscopically observed stars between the base of the giant branch and the clump, 7 stars which scatter away from the mean relation are eliminated though, as noted earlier, the anomalous giant, 7017, is now a definite member. Only three remaining member stars scatter blueward of the giant branch and two of these are binaries. The third star, 8005, is a definite astrometric, single-star member, but its radial velocity places it at 59\% membership probability. Comparison of the radial velocities from \citet{MI14} and Paper II shows virtually identical values, consistent with a lack of variability and increasing the likelihood that the velocity deviation of 8005 from the cluster mean is real.

At the turnoff region, keeping in mind that double-lined binaries are excluded, the select sample of triangles nicely illustrates the location of the binary sequence as expected for systems with two identical stars. This sequence crosses the evolving turnoff at $V$ $\sim$ 15.5. It is potentially significant to note that all but three of the numerous single-lined binaries at the turnoff lie at or above this location.

\section{Spectroscopic Abundances}
\subsection{Metallicity Estimation: ANNA}
As mentioned above, metallicity plays a major role in the estimation of the key cluster parameters of age and distance modulus, thereby impacting the specific values derived for individual stellar masses, evolutionary phases, and, more directly, equivalent width corrections for the Fe line located near Li in the determination of the Li abundances among the cooler stars. As an alternative to our EW-based spectroscopic [Fe/H] estimates and our photometric $T_{\mathrm{eff}}$ values, we have attempted to derive [Fe/H] and $T_{\mathrm{eff}}$ for each cluster member in our sample using ANNA \citep{LB17, LB18a, LB18b}, a new, flexible, Python-based code for automated stellar parameterization. ANNA utilizes a feed-forward, convolutional neural network \citep{AR02} trained on synthetic spectra, a machine-learning technique, to infer stellar parameters of interest from input spectra. Multiple tests show that ANNA is capable of producing accurate metallicity estimates with precision competitive with our EW-based analysis. Additionally, ANNA is capable of accurately inferring $T_{\mathrm{eff}}$ from our spectra alone, providing an alternate temperature determination for each star. A deep discussion of ANNA's design and capabilities can be found in \citet{LB18a}, but a short  summary of its operation can be found in \citet{AT18a}. ANNA is freely available for download; the version of ANNA used in this investigation can be found at Zenodo \citep{LB17}, while the current version of the current version of the code can be found at GitHub.\footnote{https://github.com/dleebrown/ANNA}

As a starting point for deriving the cluster metallicity and for comparing the effectiveness of ANNA relative to the traditional EW technique of Paper II, all stars classed as nonmembers and/or binaries in Table 1 were removed from the sample. This cut reduced the sample to 268 stars, higher than the 251 stars of Paper II restricted using ground-based astrometry. For future reference, the average [Fe/H] for all these stars using Paper II abundances is [Fe/H] = -0.038 $\pm$ 0.104 (sd). If, as in Paper II, the four reddest/coolest stars with $(B-V)_o \geq 1.35$ are removed, the remaining 264 stars have [Fe/H] = -0.033 $\pm$ 0.091 (sd). 

By comparison, ANNA generated a mean [Fe/H] = -0.049 $\pm$ 0.099 (sd) from 267 stars; star 13002 failed to converge to a coherent solution and was dropped from the analysis. Removal of the 4 coolest stars has a negligible impact upon the average or the dispersion. In fact, the only exclusion from the sample which has any impact on the the average is the removal of the two stars with the most deviant abundances, 12002 and 35008 at [Fe/H] = 0.67 and 0.38, respectively. For the remaining 265 stars of all colors, [Fe/H] = -0.053 $\pm$ 0.085 (sd). It should be noted that this is an improvement over the pattern identified in the metal-poor cluster, NGC 2506, where, with a limited wavelength range ($\sim$400 \AA) and $R \sim$ 13000, ANNA's reliability declined noticeably for hotter stars. The reduction in spectral features at higher $T_{\mathrm{eff}}$ minimized the sensitivity to changes in [Fe/H] and $T_{\mathrm{eff}}$; for metal-rich stars at the very cool $T_{\mathrm{eff}}$ end, the reverse issue, too rich a spectroscopic palette, can reduce the applicability of the code. 
Consisent with ANNA analysis of stars in NGC 2506 \citep{AT18a}, 
the stellar parameters generated by ANNA, [Fe/H], $T_{\mathrm{eff}}$, log $g$, and $v_t$, exhibit increasing scatter compared to values from alternative. This trend is interpreted as an indicator that the  
parameters which carry the most weight in defining the final optimal match to the observed spectra follow a similar order, i.e. the dominant parameter in constraining the neural network is the metallicity while the microturbulent velocity is the least impactful.

As illustrated in Fig. 2, the ANNA abundances show no trend with $T_{\mathrm{eff}}$. The larger scatter among the hotter stars is defined in part by a lower signal-to-noise ratio (S/N) among a few of the stars at the turnoff, coupled with the weaker line spectrum within increasing $T_{\mathrm{eff}}$. It is important to note, however, that, as shown in Fig. 3, the derived mean [Fe/H] remains independent of the S/N, as was also found for the traditional metallicity analysis of Paper II. The two anomalous points in both figures are the aforementioned 12002 and 35008. The ANNA-based [Fe/H] values for all members are listed within Table 1. 

\begin{figure}
\figurenum{2}
\plotone{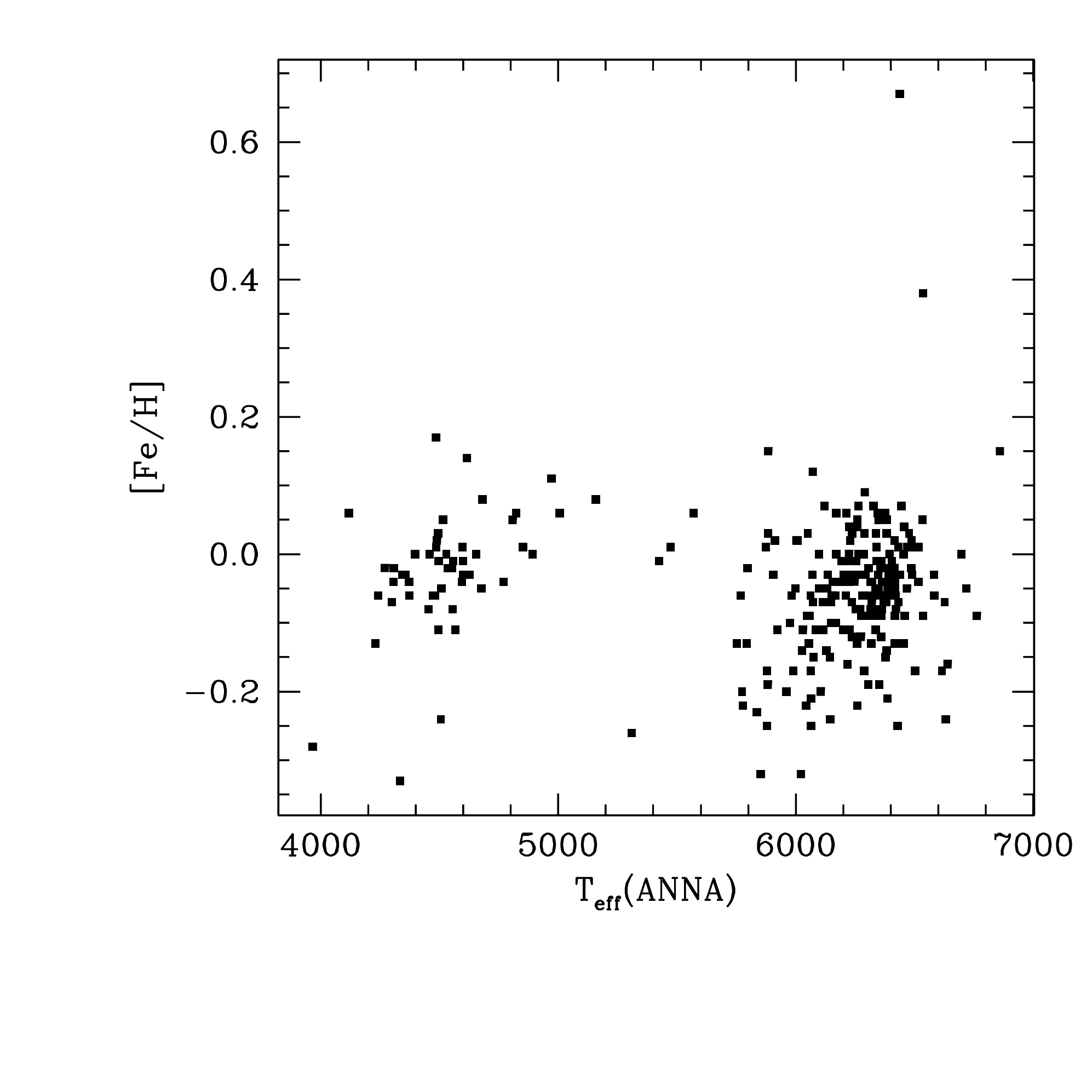}
\caption{[Fe/H] derived from ANNA for 267 stars in NGC 6819 as a function of their ANNA-derived $T_{\mathrm{eff}}$.}
\end{figure}
\begin{figure}
\figurenum{3}
\plotone{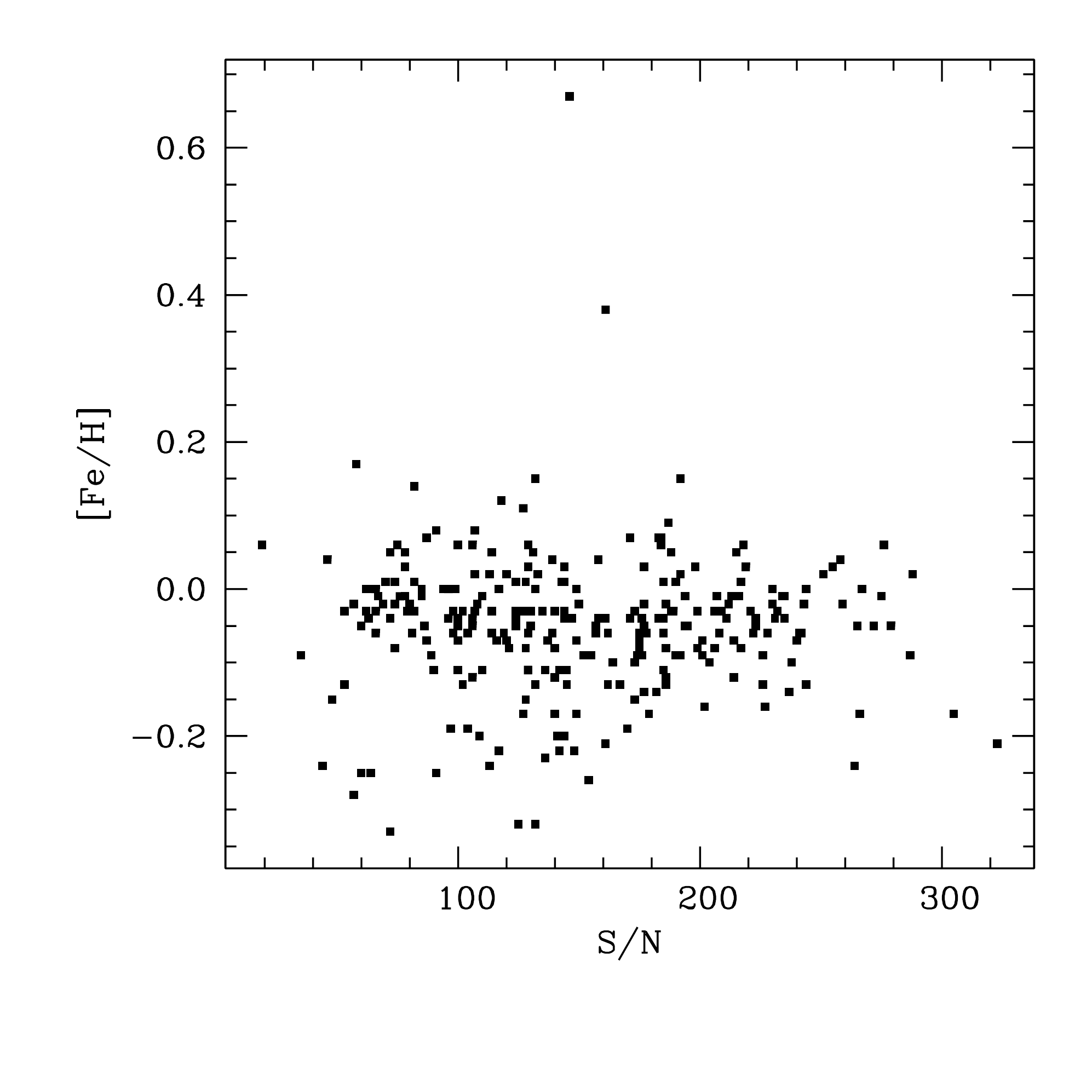}
\caption{Same as Fig. 2 as a function of the spectroscopic S/N.}
\end{figure}

\subsection{Temperature: ANNA}
The estimation of $T_{\mathrm{eff}}$ in Paper II came from reddening-corrected $(B-V)$ indices, with separate calibration relations for turnoff stars \citep{DE02} and red giants \citep{RA05}, blended to supply a smooth transition across the subgiant branch, a procedure adopted in our earlier investigations \citep{AT09, CU12}. With $T_{\mathrm{eff}}$ from ANNA, we can first check if the derived values exhibit a plausible correlation with increasing color index, noting that with the exception of added scatter due to the additional correction caused by variable reddening, the precision of the $(B-V)$ indices is high enough that the typical scatter in $T_{\mathrm{eff}}$ from photometric errors alone should be just under $\pm$80 K; for the giants, the comparable number is $\pm$50 K. Scatter caused by uncertainties in the reddening could easily double these estimates. 

Fig. 4 shows the trend of ANNA $T_{\mathrm{eff}}$ as a function of reddening-corrected $(B-V)$. The trend of decreasing $T_{\mathrm{eff}}$ with increasing $(B-V)_o$ is obvious, but the relation loses sensitivity for giants redder than $(B-V)_o$ = 1.2 or $\sim$4200 K. This is consistent with the $T_{\mathrm{eff}}$ pattern found for NGC 2506 \citep{AT18a} and is tied to the increasing complexity of the line structure in the spectra for the coolest giants over too restricted a range in bandpass. At the hotter end, there is a trend of increased $T_{\mathrm{eff}}$ with decreasing $(B-V)_o$, with an asymmetric scatter toward lower $T_{\mathrm{eff}}$ at a given $(B-V)_o$. A quick glance at Fig.1 suggests a possible explanation for the scatter. At the turnoff of the CMD, stars in the $(B-V)$ = 0.6 to 0.7 ($(B-V)_o$ = 0.44 to 0.54) range are a mixed population of moderately unevolved stars, stars within the red hook prior to the hydrogen-exhaustion-phase, and even brighter subgiants. Unlike the usual photometrically-defined $T_{\mathrm{eff}}$, where stars at the turnoff with the same color but slightly different evolutionary phases are treated identically, ANNA has the option of modifying the surface gravity and $T_{\mathrm{eff}}$ to optimize the match of the true spectrum to a synthetic analog. Unfortunately, the stars which define the low $T_{\mathrm{eff}}$ extension do not fall within a specific category of evolutionary phase at a given $(B-V)_o$. It should be noted, however, that the typical log $g$ as derived by ANNA for the excessively cool stars is lower on average by approximately 0.2 dex than that for the stars that fall upon the mean relation, despite a similar distribution in $V$.

If we exclude the 10 stars with the largest discrepancies in $T_{\mathrm{eff}}$ between the ANNA values and the photometric estimates, the mean offset, in the sense (PHOT - ANNA) is +117 $\pm$ 144 K for 257 giants and dwarfs. As an external check, perhaps the best comparison among multiple sources, at least for the cooler giants, comes from the data of \citet{HA17}, who derived $T_{\mathrm{eff}}$ from $(V-K)$, adopting a fixed value of $E(B-V)$ = 0.15 for all stars, and the color-$T_{\mathrm{eff}}$ relation of \citet{CV14}. Comparison of their $T_{\mathrm{eff}}$ to that from the $(b-y)$ of \citet{CA14} using the color-$T_{\mathrm{eff}}$ relation of \citet{RA05}, to the $T_{\mathrm{eff}}$ from \citet{CA14}, to the photometric $T_{\mathrm{eff}}$ of Paper II, and to the $T_{\mathrm{eff}}$ from APOGEE \citep{PI14} produces offsets, in the sense (HA17 - LIT), of -48 $\pm$ 51 K from 52 giants, +14 $\pm$ 35 from 52 giants, +35 $\pm$ 41 from 50 giants, and +55 $\pm$ 48 from 30 giants, respectively. It should be noted that the residuals between the \citet{HA17} and Paper II show a clear trend with $E(B-V)$ in that stars with larger $E(B-V)$ corrections show a smaller residual $T_{\mathrm{eff}}$ than stars with smaller $E(B-V)$; adjustment for this effect reduces the scatter among the residuals to $\pm$33 K.  

For the current discussion, however, the agreement among the multiple modes of deriving $T_{\mathrm{eff}}$ for the giants gives us some encouragement that the original $T_{\mathrm{eff}}$ scale of Paper II for the giants is a more reliable representation of the true system for evaluating the Li abundances. Since the Li abundance is being derived using the same EW-based approached as the [Fe/H] determination of Paper II and the ANNA abundances for Fe are in excellent agreement with those of Paper II, we will retain the color-based $T_{\mathrm{eff}}$ of Paper II for both the giants and dwarfs throughout the remainder of this Li analysis.  

\begin{figure}
\figurenum{4}
\plotone{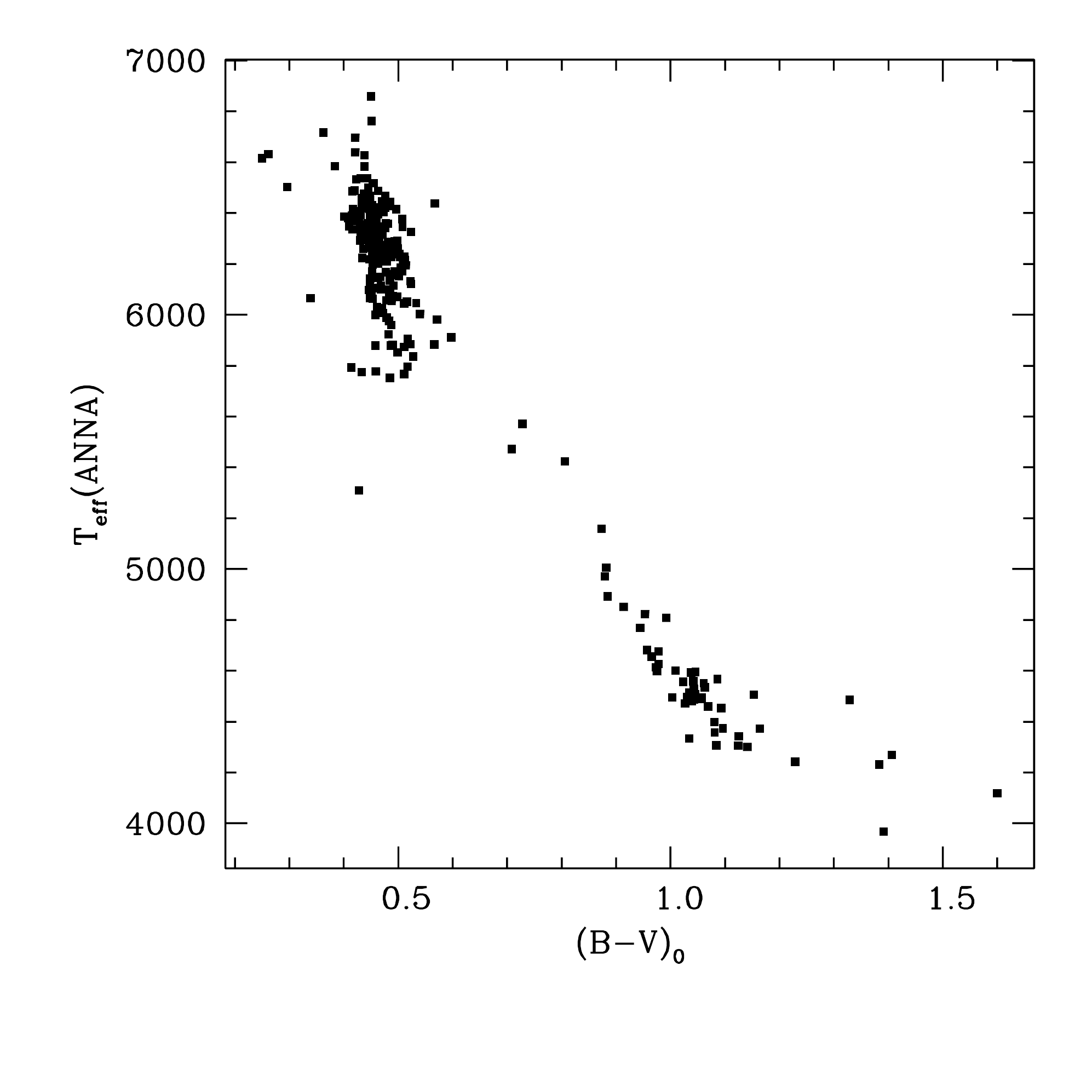}
\caption{$T_{\mathrm{eff}}$ from ANNA as a function of $(B-V)_o$}
\end{figure}

\subsection{Spectroscopic Analysis: Lithium Abundances}
We employ two methods to estimate the abundance of Li, a curve-of-growth-based computation using a direct measurement of the equivalent width of the line at 6708 \AA\ and spectrum synthesis. As the latter technique is particularly helpful for cooler stars with more blended spectra, spectrum synthesis was employed for all stars with $(B-V) \geq 0.70$, making use of the 2010 version of the MOOG software suite \citep{SN73} and a line list used and discussed in \citet{CU12}.  In spectrum synthesis, a model spectrum is constructed for each star, employing specifically chosen $T_{\mathrm{eff}}$, log $g$ and microturbulent velocity parameters as input; values for these parameters for each star are identical to those used for spectroscopic analysis described in Paper II and are found in Table 1 of that paper. Examples of spectra illustrating the region near the Li line may be found in Fig. 1 of Paper I.

For the warmer stars, equivalent widths were measured using the SPLOT utility within the IRAF spectroscopic data reduction packages \footnote{IRAF is distributed by the National Optical Astronomy Observatory, which is operated by the Association of Universities for Research in Astronomy, Inc., under cooperative agreement with the National Science Foundation.}. We use each star's temperature and the cluster iron abundance to numerically deblend the nearby Fe I line at 6707.4 \AA\ from the Li line at 6707.8 \AA\ , then use the ``corrected'' equivalent width, the star's $T_{\mathrm{eff}}$ and a grid of curve-of-growth abundances developed by \citet{ST03} from MOOG model atmospheres and employed by \citet{SD04}. The spectrograph pixel-wavelength scale, measured Gaussian full-width of the line, and the S/N per pixel are used to compute an equivalent width error for each star, utilizing a prescription originally proposed by \citet{CA88} and reformulated by \citet{DP93}. For a significant detection of Li, we require that the Li equivalent width, following subtraction of the Fe I contribution, be at least 
three times the estimated error in the equivalent width. Table 1 includes a summary of our Li abundances for all stars and the final errors in the abundance for stars with measurable Li. Photometric $T_{\mathrm{eff}}$, surface gravities, microturbulent velocities, and rotational velocities for all stars are contained in Paper II and will not be repeated.
\floattable
\begin{deluxetable}{rccccrrrrr}
\tablenum{1}
\tablecaption{Li Abundance Values for Stars in NGC 6819}
\tablewidth{0pt}
\tablehead{
\colhead{$\rm{ID_{WOCS}}$} & \colhead{$\rm{{PMga}}$} &\colhead{$\rm{PARga}$} &\colhead{$\rm{BIN}$} &\colhead{$\rm{MEM}$} &\colhead{$\rm{S/N}$} &\colhead{$\rm{T_{ANNA}}$} & 
\colhead{$\rm{[Fe/H]}$} & \colhead{$\rm{A(Li)}$} & \colhead{$\sigma_{A(Li)}$} } 
\startdata
1002 & Y & Y & N & Y & 57 & 3967 & -0.28 & 0.30 & 0.08 \\
1004 & Y & Y & N & Y & 66 & 4242 & -0.06 & -1.20 & 0.00 \\
1007 & Y & Y & N & Y & 53 & 4231 & -0.13 & -1.50 & 0.00 \\
1014 & Y & Y & N & Y & 57 & 4270 & -0.02 & -1.50 & 0.00 \\
1016 & Y & Y & N & Y & 19 & 4119 & 0.06 &  &  \\
2001 & Y & N & N & N &  &  &  &  &  \\
2003 & Y & Y & N & Y & 60 & 4676 & -0.05 & 0.60 & 0.00 \\
2004 & Y & Y & N & Y & 100 & 4372 & -0.04 & -1.00 & 0.00 \\
2006 & Y & Y & N & Y & 72 & 4594 & -0.04 & 0.35 & 0.00 \\
2007 & Y & Y & N & Y & 76 & 4558 & -0.01 & 0.35 & 0.00 \\
     &   &   &   &   &    &      &        &     &      \\
2012 & Y & Y & SB1 & Y &  &  &  & -0.50 & 0.00 \\
2016 & Y & N & N & N &  &  &  &  &  \\
3001 & Y & Y & N & Y & 129 & 4496 & -0.11 & 0.35 & 0.00 \\
3003 & Y & Y & N & Y & 44 & 4506 & -0.24 & 0.00 & 0.00 \\
3004 & Y & Y & N & Y & 113 & 4489 & 0.02 & 0.35 & 0.00 \\
3005 & Y & Y & N & Y & 63 & 4459 & 0.00 & 0.35 & 0.00 \\
3007 & Y & Y & N & Y & 74 & 4596 & 0.01 & 0.35 & 0.00 \\
3009 & Y & Y & N & Y & 139 & 4481 & -0.06 & 0.35 & 0.00 \\
3021 & Y & Y & N & Y & 98 & 4598 & -0.03 & 0.30 & 0.00 \\
4001 & Y & Y & N & Y & 85 & 4497 & -0.01 & 0.30 & 0.00 \\
\enddata
\end{deluxetable}

\section{Li Patterns}
Derived values of A(Li) are shown in Figs. 5 through 10 for stars other than those designated as nonmembers. 
We now discuss the apparent trends among the stars as a function of their evolutionary phase.
\subsection{Li: The Turnoff Region}
Fig. 5 shows the trend of Li with reddening-corrected $V$-magnitude for all stars bluer than $(B-V)_o$ = 0.54 in Fig. 1. Fig. 6 illustrates the same sample plotted as a function of $(B-V)_0$. The adopted mean reddening for these two figures is $E(B-V)$ = 0.16 and $A_V$ = 0.50.

\begin{figure}
\figurenum{5}
\plotone{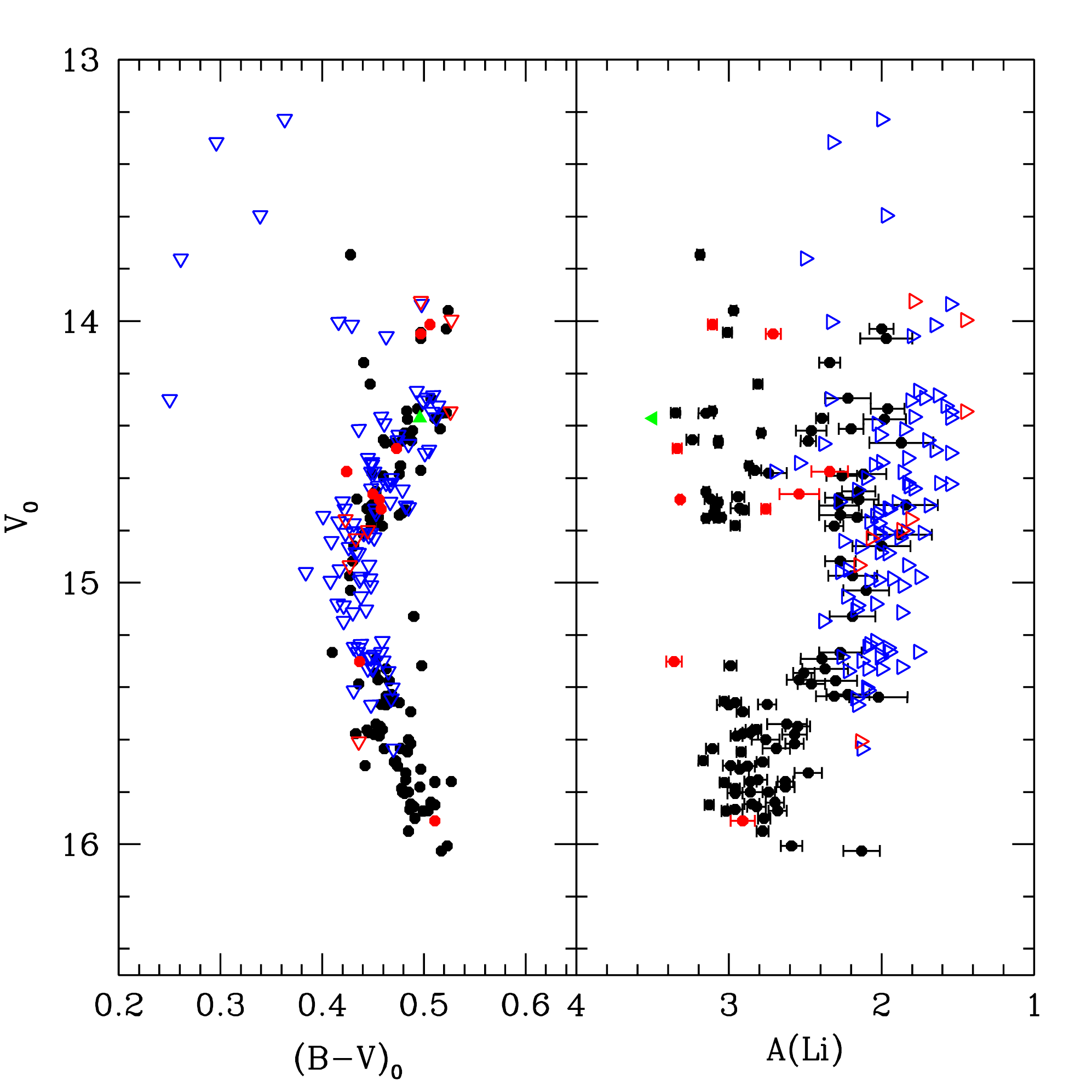}
\caption{A(Li) for stars bluer than $(B-V)_o = 0.54$ as a function of $V_0$. Triangles are stars with only upper limits to A(Li) while solid circles with error bars identify detections. Red symbols denote single-lined binaries while blue and black denote single stars.  For 18019 the Li measure implies a lower limit and is plotted as a filled green triangle. The location of the main sequence Li dip is clearly visible, centered at $V_0 \sim 15.1$.}
\end{figure}

\begin{figure}
\figurenum{6}
\plotone{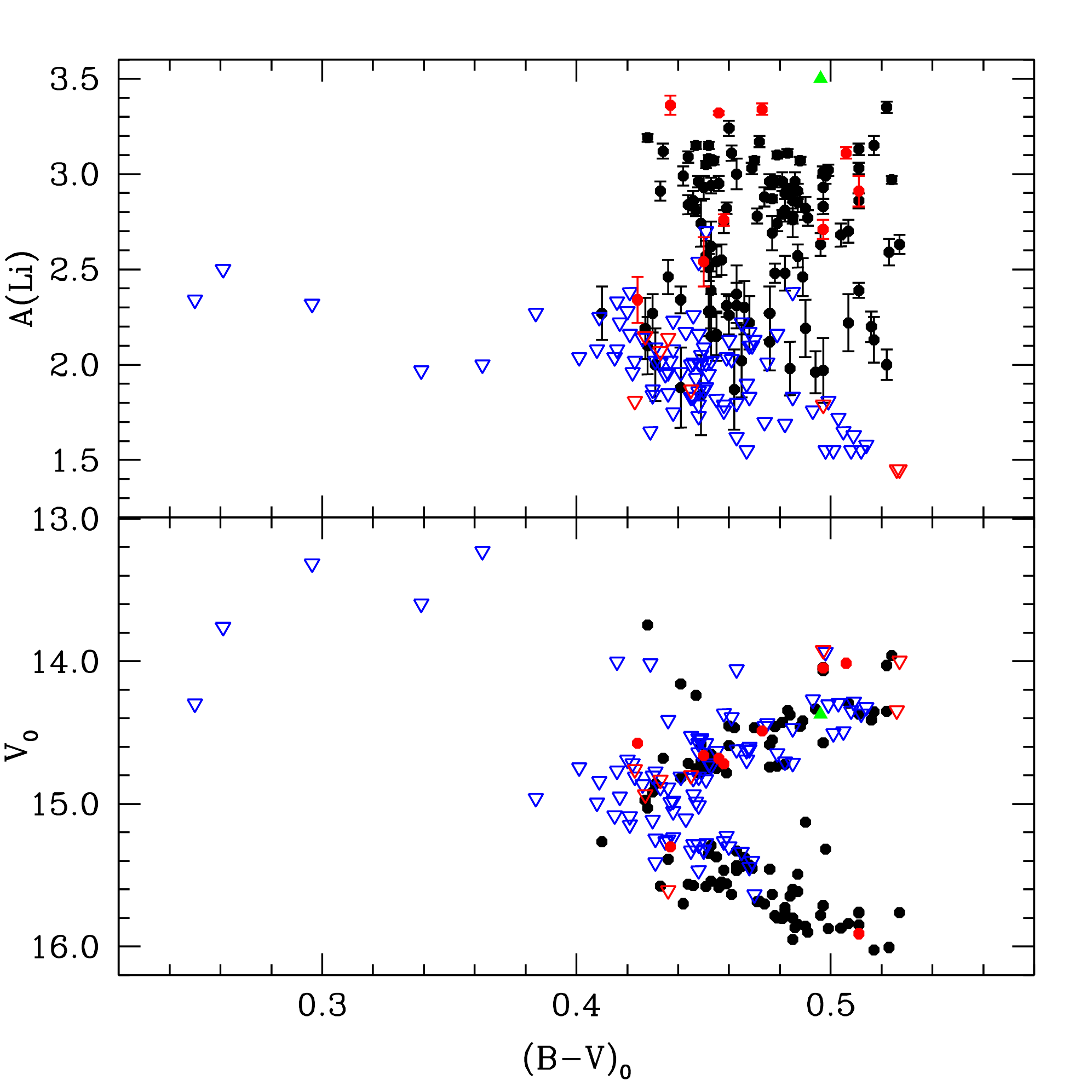}
\caption{A(Li) for turnoff stars as a function of $(B-V)_0$. Symbols have the same meaning as Fig. 5.}
\end{figure}

The first feature of importance is the limiting A(Li) for stars on the brighter or higher mass side of the Li dip. All single-star members have A(Li) = 3.35 or less. Of the six stars with A(Li) above 3.2, only two are single and the remaining four are classed as single-lined binaries (red and green points). The five stars that are situated both brighter and bluer than the cluster isochrone turnoff, supposedly blue stragglers of potentially binary origin, all exhibit upper limits to A(Li) of 2.5 or less.  One star, 10010, with measurable A(Li) at 3.19 is located in the CMD at the tip of the isochrone blue hook for single stars about to enter the subgiant branch.  

A key property which makes NGC 6819 invaluable for probing the nature of the Li dip is its age. While older clusters such as NGC 6253, M67, and NGC 188 \citep{CU12, PA12, RA03} have been studied in the mass range populating the Li dip, they are too old for mapping the high-mass edge of the distribution. The stars at the top of the turnoff feeding the subgiant and giant branches in these clusters come from the Li dip itself, making the exact boundary impossible to define and ensuring that the giants exhibit no significant Li abundance, even before convection and mixing kick in. The value of the stars more massive than the Li dip stems from the often-made assumption and prediction from SSET that if any stars within a cluster retain the signature of the primordial cluster abundance, it should be these stars. However, near 7000 K, upwards diffusion may be enriching the surfaces of slower rotators \citep{RM93}, as evidenced by the super-Li-rich dwarf J37 in the Hyades-aged cluster, NGC 6633 \citep{DE02}. Among the clusters of intermediate age studied to date, NGC 752, NGC 3680, IC 4651 \citep{AT09}, NGC 2506 \citep{AT18a}, NGC 7789 \citep[][in prep.]{AT19}, and now NGC 6819, there is evidence in each cluster for a limiting A(Li) value typically between A(Li) = 3.2 and 3.35, consistent with the primordial solar system value of A(Li) = 3.30 \citep{AG89}. However, it is also true that every cluster exhibits a range of A(Li) which often extends to 2.8 or lower. Since the normal stars within the hydrogen exhaustion phase at the top of the turnoff and beyond have evolved off the main sequence en route to their current locations, it is perhaps unsurprising that some of the brighter turnoff stars have A(Li) well below their cluster limit. What is surprising, however, is the changing fraction of turnoff stars which fall below the given cluster limit, depending upon the age. For NGC 6819, among stars in the red hook at essentially identical magnitudes and supposedly similar evolutionary phase, A(Li) can range from the detection limit of $\sim$ 3.2 to an upper limit of less than 1.6. We will return to this issue in Section 6.

Moving down the turnoff toward fainter $V$, the next striking feature is the sharp transition from detectable A(Li) near 3.2 to stars with detections or upper limits below 2.3. This edge occurs over a magnitude range of less than 0.1 mag near $V_0$ $\sim$ 14.75, a range comparable to the combined photometric and reddening uncertainties alone. The Li dip remains deep to $V_0$ $\sim$ 15.25, where a more gradual rise in detectable A(Li) begins, plateauing to a fixed value near $V_0$  $\sim$ 15.45. For stars fainter than this edge, the degree and range of evolution off the ZAMS should be significantly less than for the stars on the bright side of the Li dip. Despite this, if we bin the stars fainter than the dip by $V_0$ using bins 0.1 mag wide between $V_0$ = 15.50 and 15.90, excluding binaries and upper-limits, the mean A(Li) for the four bins is 2.83, with a dispersion among the bin averages of only 0.05 dex. By contrast, the dispersion within each bin ranges from 0.15 to 0.22 dex, with a dispersion among all stars between 15.5 and 15.9 of 0.17 dex; the predicted dispersion from the spectroscopic errors is $\pm$0.05 dex. We conclude that the abundance scatter among the stars on the less evolved portion of the main sequence is real.

If the Li dip profile in $V$ is symmetric \citep{CU17}, the center is located near $V_0$ = 15.1 $\pm$ 0.1 mag or $M_V$ = 3.2 for $E(B-V)$ = 0.16; the analogous numbers for $E(B-V)$ = 0.14 and 0.12 are $M_V$ = 3.3 and 3.4, respectively. Determination of the exact profile of the dip is challenging since the majority of Li measures at its center are defined only by upper limits and the rise to a Li plateau among the lower mass stars appears more gradual than the sharp boundary among the higher mass stars.  Using the isochrones of Fig. 1, the mass at the center of the symmetric dip is 1.348 $\pm$ 0.025 $M_{\sun}$; for the lower reddening values the masses are 1.317 and 1.288 $M_{\sun}$, respectively.  Under the assumption that the profile of the Li dip in NGC 6819 has the same shape as the more asymmetric curve found in the Hyades and Praesepe by \citet{AT09} evolved to the age of NGC 6819, the A(Li) minimum shifts to a mass of 1.36 $\pm$ 0.02 $M_{\sun}$.  A mapping of the reddening-corrected $V$ magnitude to initial mass for $E(B-V)$ = 0.16 from the VR isochrones is illustrated in Fig. 7; we emphasize that the mapping of mass to magnitude is virtually indistinguishable if the $Y^2$ isochrones are used. The $Y^2$ isochrones generate masses typically larger by 0.008 $M_{\sun}$. The distribution is cut off at the red edge of the hydrogen-exhaustion phase since the spread in mass among stars on the subgiant branch through the tip of the giant branch is less than 0.03 $M_{\sun}$ and would crowd all the evolved stars into a single narrow vertical band at the right of the figure.

\begin{figure}
\figurenum{7}
\plotone{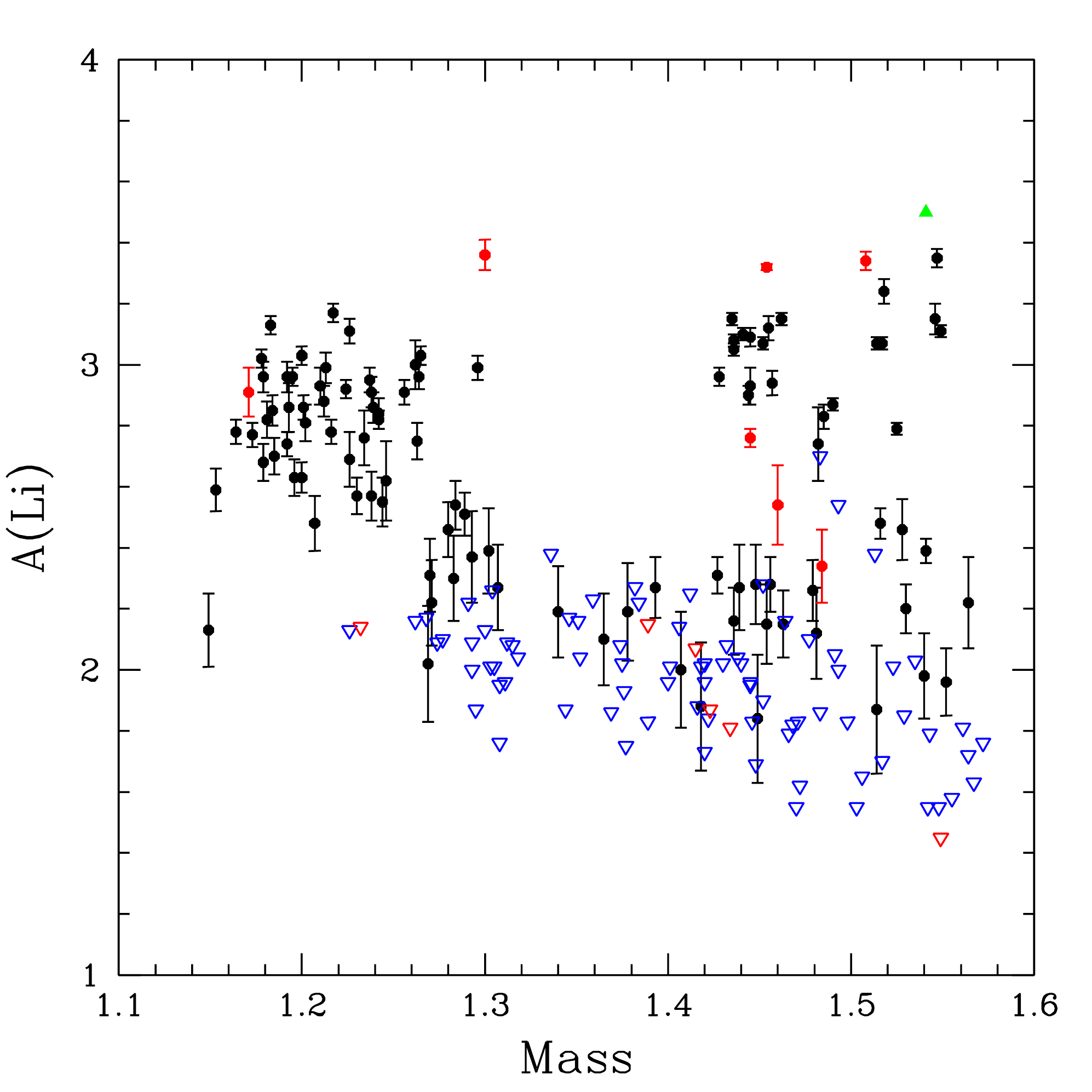}
\caption{Lithium abundances for main sequence and turnoff members through the red hook, plotted as a function of initial mass based on the 2.25-Gyr isochrone comparison displayed in Fig. 1. Symbol colors and types are as in Fig. 5.}
\end{figure}

The Li dip, central-mass-$[Fe/H]$ relation derived by \citet{AT09} from several open clusters and the $Y^2$ isochrones predicts that for $[Fe/H] = -0.04 \pm 0.03$, the mass for the Li dip center should be $1.36 \pm 0.04$ $M_{\sun}$, entirely consistent with results displayed in Fig. 7 after adjustment for the small mass offset between $Y^2$ and VR. For $[Fe/H] = +0.09$, the predicted central mass is 1.42 $M_{\sun}$, essentially the same as found in the Hyades. 

\citet{CU12} present a different relationship between $[Fe/H]$ and Li dip mass, relevant for older clusters for which stars on the hotter side of the dip are no longer present on the main sequence. This alternative relation pegs the masses of stars in the Li plateau on the cool side of the Li dip, with a similar dependence on metallicity; for the lower abundance cited above, stars on the cool side of the Li dip in NGC 6819 should have initial masses at 1.20 $M_{\sun}$, in excellent agreement with the initial mass associated with the peak A(Li) value on the low-mass side of the main sequence Li dip.

Two critical independent checks on the validity of the mass scale, independent of the Li dip, are currently available. First, one can make use of the mass estimates for three stars in the eclipsing binaries analyzed by \citet{BR16} which overlap in apparent magnitude with the data in Figs. 1 and 5. For eclipsing binary stars 23009A, 24009A, and 40007A, \citet{BR16} derive $V$ = 15.13, 15.74, and 16.11, respectively. Neglecting the small effect due to the possible variation in reddening, we can translate these stars to the appropriate location in the CMD and derive their individual masses from the same isochrones used to construct the mass trend in Fig. 7. The isochronal masses are 1.23, 1.31, and 1.47 M$_{\sun}$, while the binary mass determinations are 1.218 $\pm$ 0.008, 1.251 $\pm$ 0.057, and 1.464 $\pm$ 0.011 M$_{\sun}$, respectively. 

Second, building upon the asteroseismological data for red giants in NGC 6819, \citet{HA17} derive masses for first-ascent red giants and red clump stars, obtaining 1.61 $\pm$ 0.02 and 1.64 $\pm$ 0.02  M$_{\sun}$, respectively. While the VR isochrones do not include clump stars, we can assign the mass at the base of the vertical red giant branch as typical of the red giants and the stars at the tip of the RGB to have masses similar to the red clump stars. For $E(B-V)$ = 0.16 and an age of 2.25 Gyr, the red giants and clump stars have isochronal masses of 1.609 and 1.633  M$_{\sun}$. By contrast, if we lower the reddening to $E(B-V)$ = 0.14 and 0.12, with corresponding shifts in age and distance, the paired masses become  1.574-1.598  M$_{\sun}$ and 1.542-1.566  M$_{\sun}$, respectively. 

A more specific question beyond the typical mass of the stars populating the Li dip is the actual profile of the feature. Does the range of stars within the Li dip evolve over time, i.e. do the boundaries of the dip expand over time, encroaching on stars of higher and lower mass or temperature than found at the boundaries of the dip at an earlier age? To test this possibility, we make use of the Li dip profile defined by the Hyades and Praesepe clusters as discussed in \citet{AT09} and revised by \citet{CU17}. To minimize the impact of metallicity, we transfer the $(B-V)$-based relation illustrated in Fig. 7 of that paper to a $T_{\mathrm{eff}}$-based relation, building upon the long-standing observation that the physical mechanism controlling the Li dip is solely temperature-dependent \citep{BA95, CH01, AT09, CU12, RA12a}, explaining why higher metallicity stars in the Li dip have higher masses. For internal consistency, we have converted the A(Li) vs $(B-V)_0$ to A(Li) vs $T_{\mathrm{eff}}$ using the color-temperature relation from a VR isochrone of age 0.6 Gyrs and [Fe/H] = +0.13, the isochrone set with a metallicity closest to that derived for the Hyades and Praesepe \citep{CU17}, [Fe/H] = +0.15. Using an isochrone of any age between 0.3 Gyr and 0.9 Gyr leaves the conclusions unchanged.

Fig. 8 shows the Hyades/Praesepe data with no Li upper limits or binaries included, superposed on the A(Li) trend with $T_{\mathrm{eff}}$ for NGC 6819 at the age of the Hyades 
, i.e. the masses of the stars occupying the vertical turnoff of NGC 6819 have been used to derive their $T_{\mathrm{eff}}$ at the time the cluster had the same approximate age as the Hyades.
The $T_{\mathrm{eff}}$ scale for the A(Li) profile has been adjusted by adding 150 K to the temperatures 
to align the Li-dip with that of the Hyades; use of the $Y^2$ isochrones would lead to a smaller shift of 120 K. The need for the shift can have multiple origins tied to the theoretical evolutionary rates for stars of varying mass as predicted by the isochrones, the $(B-V)_0$ - $T_{\mathrm{eff}}$ conversion relation, and the adopted reddening and distance modulus. The temperature adjustment could imply that the masses of the stars populating the vertical turnoff in NGC 6819 are too large compared to the unevolved main sequence stars by about 0.05 $M_{\sun}$ and/or the adopted distance modulus which defines $M_V$ for stars in the Li dip is too large.  As noted above, a reduction in $E(B-V)$ from 0.16 to $\sim0.13$ would produce the appropriate change in the derived stellar masses. However, the related changes in both the distance modulus and the masses are contradicted by the excellent agreement with the eclipsing binary analysis \citep{BR16} and asteroseismology \citep{HA17}. 

\begin{figure}
\figurenum{8}
\plotone{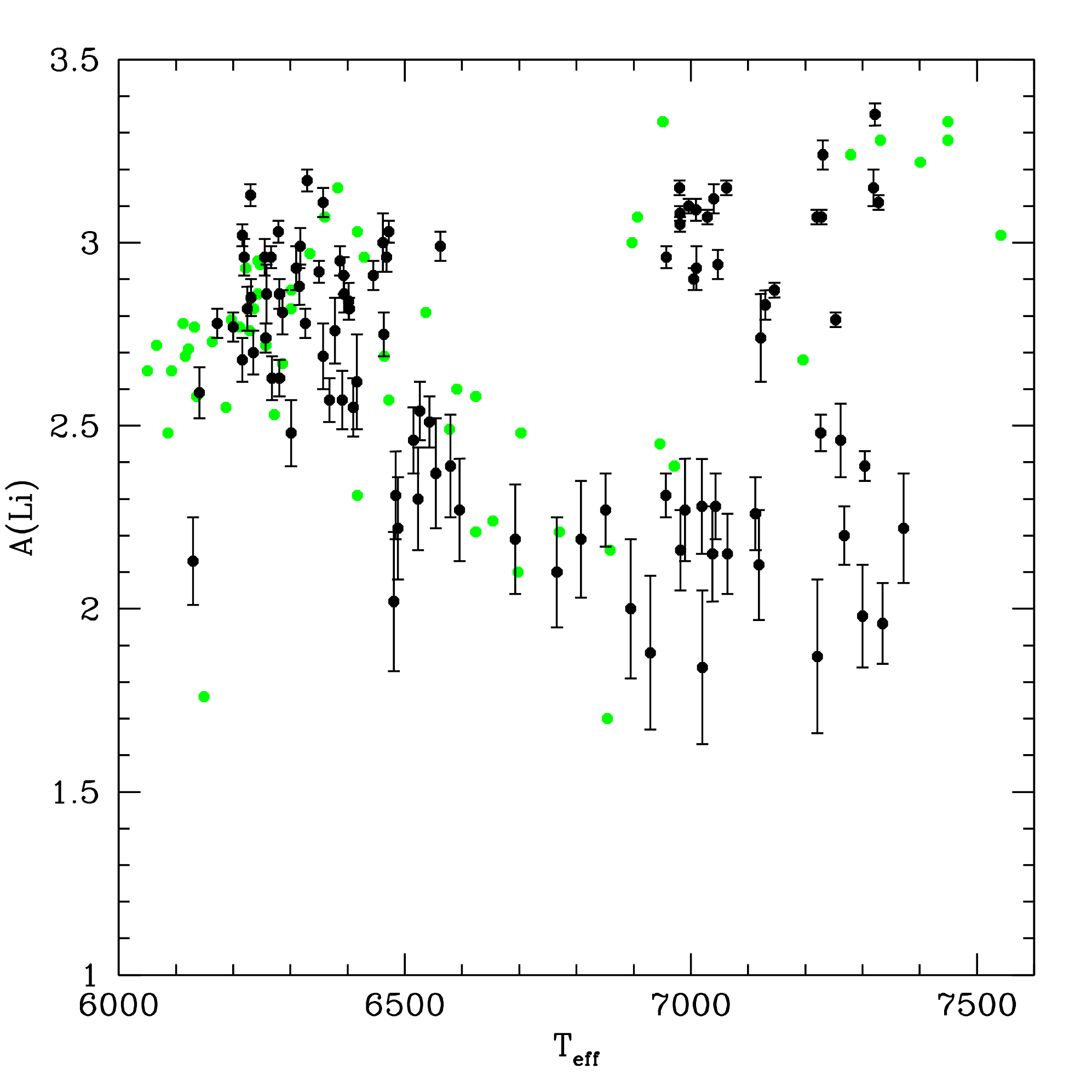}
\caption{A(Li) for stars at the turnoff of NGC 6819 as a function of their main sequence temperature at 0.6 Gyr using VR isochrones.
Symbols have the same meanings as in Fig. 5. Green circles are the Hyades/Praesepe Li measures of \citet{CU17} with isochronal $T_{\mathrm{eff}}$ shifted by +150 K.}
\end{figure}

What is striking about the comparison is the virtually identical $T_{\mathrm{eff}}$ range in the two profiles. The most Li-rich stars at the cool and the hot edges of the Li dip sit on or just outside the Hyades/Praesepe profile. At the hot edge, as already noted, the A(Li) limit based upon a pair of stars in NGC 6819 is as high as that found in the younger clusters (A(Li) $\sim$ 3.35), but the majority of stars scatter toward lower A(Li), unlike the minimal spread at a given $T_{\mathrm{eff}}$ in the combined Hyades/Praesepe sample. At the cool edge, only one star in NGC 6819 sits within the Li dip, i.e. systematically above the band defined by the younger clusters. In the mean, the stars between 6400 K and 6600 K fall on or below the trend defined by the younger clusters, potentially indicating that the Li-depletion mechanism in this temperature regime continued to reduce the surface abundance beyond the value predicted at 0.65 Gyr. The viability of this claim ultimately depends upon the assumed difference in the initial cluster A(Li) for the comparison clusters with significantly different metallicities. One could shift the Li data for the Hyades/Praesepe sample down by 0.2 dex to remove the separation between the two samples but this would place the coolest stars beyond the dip in the older cluster systematically above the younger stars.

The consistency of the boundaries of the Li dip when comparing NGC 6819 and the Hyades/Praesepe data, particularly at the hot edge, is important because taken individually, the statistical samples defining these rapid transitions in A(Li), especially for Hyades/Praesepe, are modest, at best.  \citet{CU17} referred to the hot edge as ``the wall" but cautioned against reading too much into a trend defined by a handful of stars. However, as discussed in Section 6, the cumulative sample afforded by the merger of data from Hyades/Praesepe \citep{CU17}, NGC 752, NGC 3680, IC 4651 \citep{AT09} and now NGC 6819, leaves little doubt that the transition from Li-rich to Li-poor among the hotter stars occurs over a very small range in mass.
\subsection{Li: Subgiants and Giants}
Figs. 9 and 10 show the run of Li abundances across the subgiant and giant branches as a function of $V_0$ and $(B-V)_0$, respectively. Because of its unique status, star 7017 is included in all cases by a magenta starred symbol.

\begin{figure}
\figurenum{9}
\includegraphics[angle=270,width=\columnwidth]{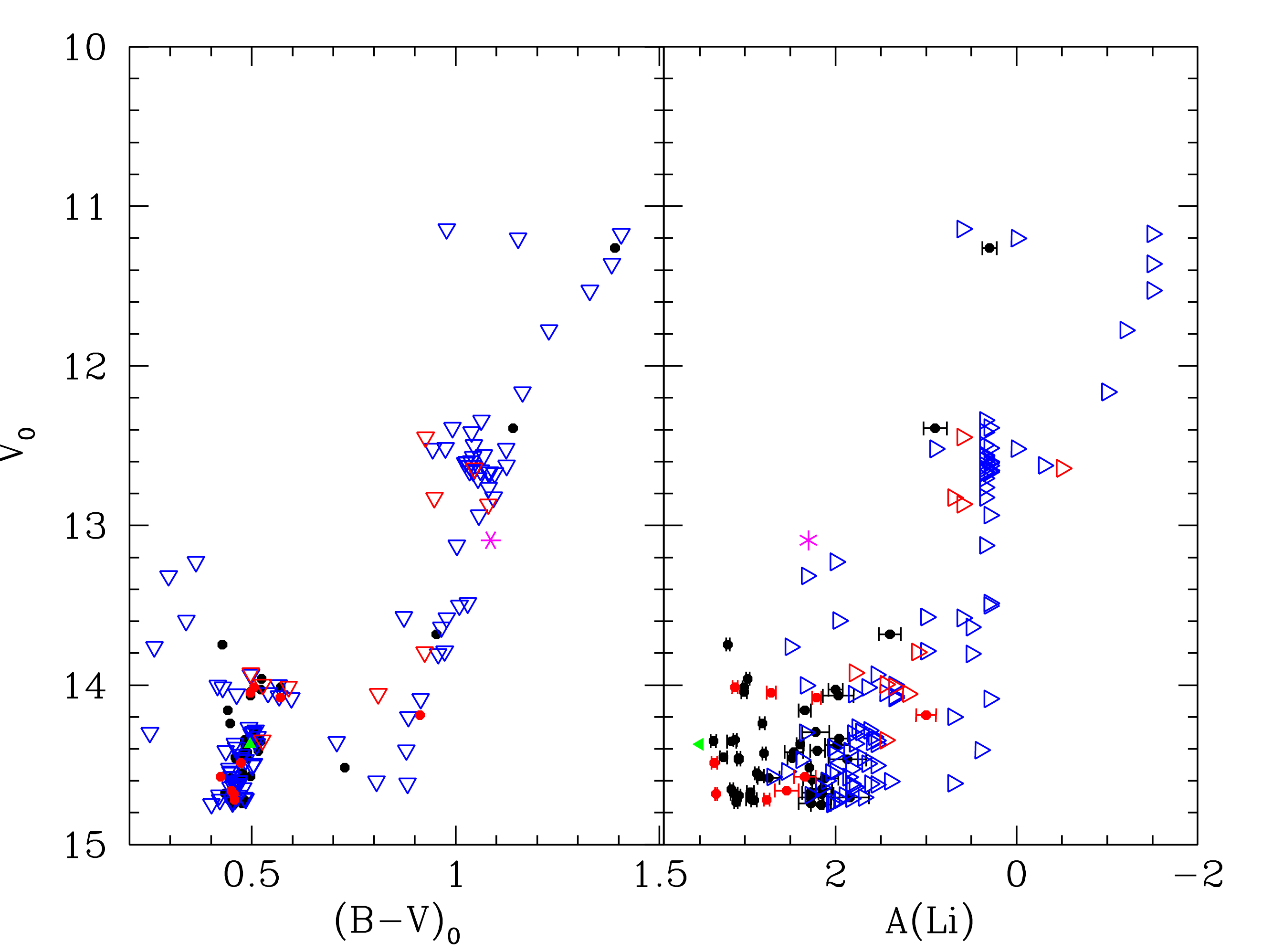}
\caption{A(Li) as a function of $V_0$ for stars brighter than $V_0 = 14.75$. Symbol colors and types are same as in Fig. 2. 7017 is shown as a magenta star.}
\end{figure}

\begin{figure}
\figurenum{10}
\includegraphics[angle=270,width=\columnwidth]{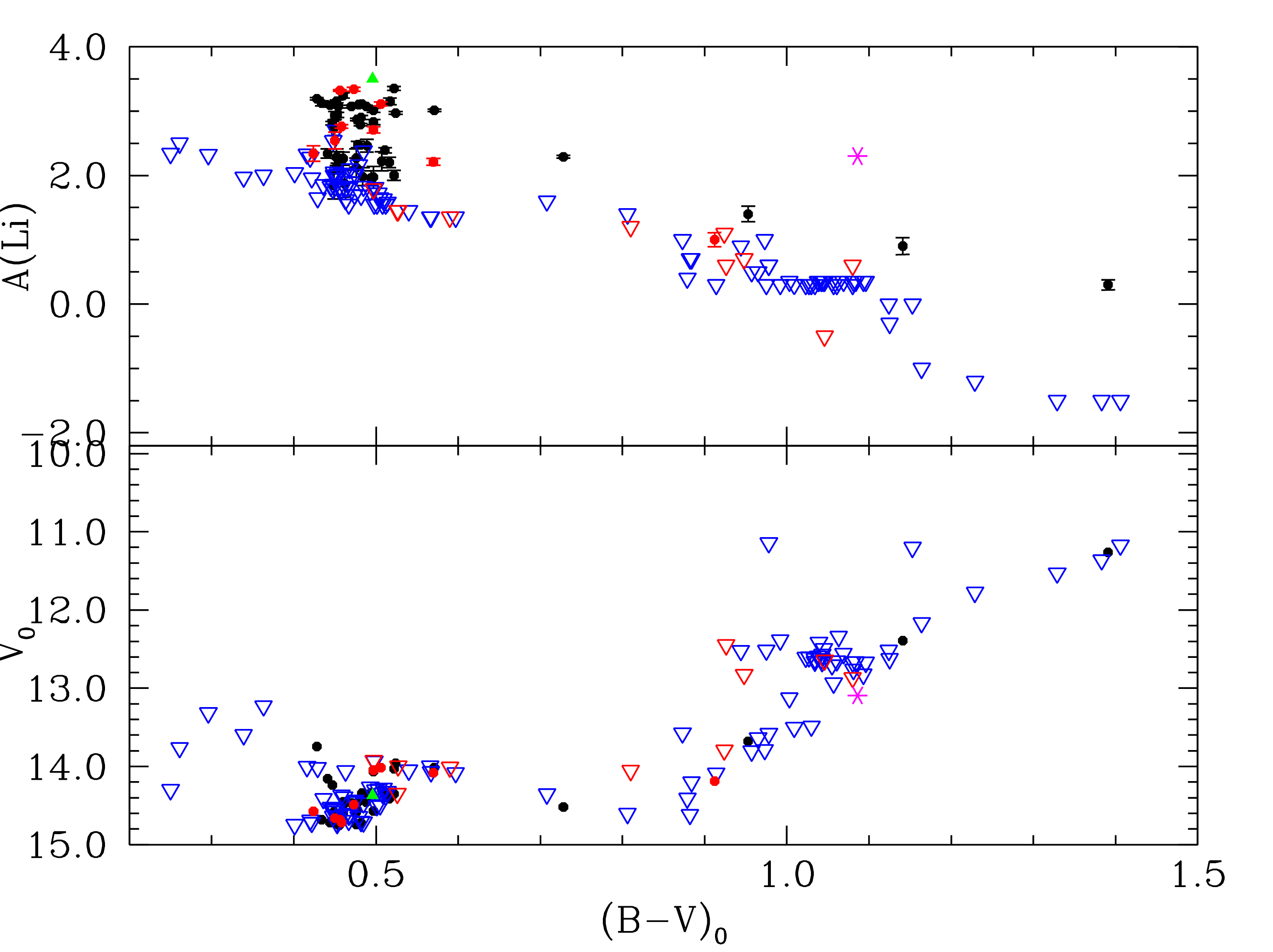}
\caption{A(Li) as a function of $(B-V)_0$ for stars brighter than $V_0 = 14.75$. Symbols have the same meaning as in Fig. 9}
\end{figure}

Choosing $(B-V)_0$ = 0.54, the coolest point on the red hook at the turnoff, as the color boundary of the subgiant branch leading to the vertical giant branch, the trend from the subgiant branch and beyond is apparent, but limited. The limitation is imposed by the fact that only upper limits to A(Li) are available for the majority of the stars. Limits determined through spectrum synthesis can be set to lower values at cooler temperatures, so the trend defined in Fig. 10 does not supply insight into how deep the Li depletion goes and at what rate for most stars leaving the main sequence. However, even at the start of the subgiant evolution, of the six stars between $(B-V)_0$ = 0.55 and 0.6, only two have detectable Li with A(Li) near 3.0 for the single star and below 2.5 for the binary. The remaining four, including one binary, already have upper limits below A(Li) = 1.4, 40 times lower than the detections still possible at the turnoff. Beyond this phase, all upper limits lie below 1.4, with the exception of one star near 1.6. Of the 65 stars populating the post-turnoff phase ($(B-V)_0$ $>$ 0.54), 8, including two binaries, have detectable Li; all stars with detectable Li on the vertical giant branch, with the exception of 7017,  have A(Li) at 1.4 or less.	 

\section{Li Evolution: Comparison of Theory and Observation}
Throughout the entire range of stellar masses where Li is observed, evidence points to rotation as the dominant non-standard mechanism affecting the surface A(Li). In most cases it is through the relationship between angular momentum loss and mixing, which is shown here for the first time to be relevant for stars more massive than those in the F-dwarf Li dip, although in young G/K dwarfs it can also be through structural and related effects due to rapid rotation.  We relate our findings from NGC 6819 to most of this mass range
\subsection{Li Dip: Origins}
To decipher the internal structure of the lower mass stars currently populating the vertical turnoff of NGC 6819 and beyond, no feature is more important than the Li dip, particularly its boundaries. A key to the nature and origin of the Li dip centers on a critical question: what stellar parameter(s) determines the edge on the hot (high mass) side of the profile? The order-of-magnitude decline in Li across the hot edge of the dip within the Hyades/Praesepe sample \citep{AT09, CU17} occurs over a color range of $\sim$0.02 mag in $(B-V)_0$, equivalent to a change in $T_{\mathrm{eff}}$ of $\sim$80 K. For the intermediate-age (1.45 Gyr to 1.75 Gyr) composite sample of NGC 752, IC 4651, and NGC 3680, the transition in the vertical turnoff occurs over a range in $V$ of less than $\sim$0.1 mag \citep{AT09}, the same as NGC 6819. Equally important, the physical mechanisms controlling dip evolution on the cool edge may be distinctly different from those among the hotter dwarfs. As illustrated in Fig. 8, the stars on the cool side of the Li dip in NGC 6819 may have undergone greater depletion over this time interval but, within the uncertainties, the width of the Li-profile remains relatively static between 0.65 Gyr and 2.25 Gyr. The uncertainties include the limitation that the majority of A(Li) measures at the centers of the Li dips in multiple clusters are only upper limits and the fact that the translation of the A(Li) profile from the evolved turnoff region of the older clusters to the comparable age of the Hyades/Praesepe sample requires an offset in the $T_{\mathrm{eff}}$ scale to bring the A(Li) profiles into alignment. 

Since its delineation by \citet{BT86}, following on earlier work by \citet{WA65}, explanations for the origin of the Li dip have focused on multiple options: mass loss exposing deeper, Li-diluted layers for stars shedding their outer atmospheres \citep{SC90, SW92}, diffusion which allows Li atoms to settle below the visible surface without the need to reach layers hot enough for its destruction \citep{MI86, RM93, CH95}, and some form of mixing or circulation mechanism, driven by gravity waves \citep{GA91, MA96, TC03, TC05}, rotation \citep{CM88, DE93, CT99, PI10}, or some combination thereof that takes the atmospheric Li to interior layers where the temperature crosses 2.5 x 10$^6$ K, Li is destroyed, and the Li-depleted material can then be dredged back up to the surface, usually via convection, meridional circulation or other mixing mechanism. The Yale rotational models \citep{ PI89, PI90}, for example, show that stars deplete surface Li as a result of mixing induced by the loss of angular momentum and the resulting instabilities that are triggered in the stellar interior. 

Of these options, the mass loss rates required to produce the level of depletion found in the sun or within the Li dip are excessive, if not  implausible, while failing to explain the apparent correlation among lithium, beryllium, and/or boron depletion for hotter stars, 
where beryllium and boron survive to progressively greater depth \citep{SW92, DE97, RA07, BO05, BO16}. 
Some evidence for diffusion comes from chemically peculiar A dwarfs \citep{RM09}; some of these show Li depletion \citep{BU97, BU98, BU00}.  However, for cooler stars within the F-dwarf Li dip, below the surface convection zones, diffusion timescales into deeper layers are longer, 
resulting in an increasing Li abundance as a function of depth. This results in Li enhancement as the stars leave the main sequence and evolve across the subgiant branch when convection mixes deeper layers back to the observable atmosphere. This pattern has not been seen for Li in M67 \citep{PI88, BA95, PA12}, with the possible exception of one star \citep{SD00}, or NGC 6253 \citep{CU12}, clusters old enough that the subgiant stars come from the mass range which defines the Li dip among unevolved main sequence stars. 
The large changes in A(Li) seen in the subgiants of these clusters seem to be caused exclusively or primarily by rotational mixing.  Similarly, no more massive, post-turnoff stars in NGC 6819 show evidence of diffusion. 
Li aside, small and subtle diffusion effects have been claimed in M67 for a range of other metals through comparison of abundances for stars below, at, and above the turnoff region \citep{BE18, GA18, SO19} with elemental differentials ranging from $\sim0.05$ to 0.3 dex.

\subsection{Li Evolution: the Role of Stellar Rotation and the Cooler Dwarfs}
\citet{KR67} showed that the distribution of rotational velocities for dwarfs hotter than mid-F drops by an order of magnitude for dwarfs cooler than mid-F. \citet{BO87} noted that the minimum in A(Li) in the Hyades coincides in $T_{\mathrm{eff}}$ with this break in the Kraft curve. The link between the drop in $V_{ROT}$ and the center of the Li dip has been confirmed in NGC 2516, M34, and NGC 6633 \citep{TE02, JE02}, among other clusters, and redefined more precisely for the Hyades by \citet{BO16} and  the combined Hyades/Praesepe sample by \citet{CU17}. As illustrated in Fig. 15 of \citet{CU17}, the minimum in the depth of the Li dip occurs at $T_{\mathrm{eff}}$ just below 6700 K, where the $V_{ROT}$ typically approaches 60 km s$^{-1}$. $V_{ROT}$ then declines in linear fashion with declining $T_{\mathrm{eff}}$ to below 10 km s$^{-1}$ near $T_{\mathrm{eff}}$ = 6200 K, where A(Li) returns to a value between 3.1 and 3.2 at the cool edge of the Li dip.

A brief outline of current understanding of the evolution of the rotation-$T_{\mathrm{eff}}$ relation is relevant.  Rotation periods can be determined from measurements of chromospheric activity (see, e.g. \citet{NO84} and \citet{SO91}), and increasingly through the photometric variation of stars caused by nonuniform surface flux, specifically star spots, as seen through satellite surveys like {\it CoRoT} \citep{BA06}, {\it Kepler} \citep{BO10}, and, most recently, {\it Gaia} DR2 \citep{LA18a}. 
Periods do not suffer from the lack of information about inclination angle ({\it sin i}) inherent in measures of line broadening \citep{AF12, ME11b, MC13a, MC13b, MC14, RE15}. {\it The broad pattern that has emerged from decades of observational analysis is a period-mass-age relation.} In clusters below 100 Myr in age, the relation is bimodal, with short-period and long-period rotation rates in the same cluster, though this bifurcation is less distinct for stars with masses below 0.4 $M_{\sun}$ and/or younger than 10 Myr \citep{BO14}. As clusters age, the long-period sequence evolves to even longer rotation periods, while the short-period sequence becomes subsumed within the long-period pattern by the age of the Hyades \citep{TE02, BA03, ME09, ME11a}. Exposing the physical processes, particularly angular momentum loss, magnetic field structure, and convection, underlying the time evolution of rotation as a function of stellar mass has been the ongoing goal of many studies \citep[e.g.][]{BA10, CR11, RE12, BR14, ST16, GR18} since the determination of an empirical power-law relation by \citet{SK72} for G dwarfs.

A large degree of scatter in A(Li) has been observed in G/K dwarfs of the Pleiades (roughly 100 Myr-old) \citep{BU87, SO93, BO18}, and also in the slightly older cluster, M35 \citep{AT18b}. 
A(Li) is correlated with $V_{ROT}$, in that stars that have retained A(Li) closer to their cluster primordial value have a significantly higher probability of falling within the bifurcated short-period category or in transition to the long-period track, i.e. they have yet to spin down to the long-period rotation rate. 

Evidence suggests rapid rotation leads to radius inflation which leads to less depletion of Li \citep{JDJ18}.  
It is unclear how this scatter at a given $T_{\mathrm{eff}}$ evolves. For example, the Hyades/Praesepe G dwarf sample of \cite{CU17} shows little scatter except near 6000K and 5200K, while the K dwarf sample (where the effect is most pronounced in the Pleiades) shows only upper limits.  Moving to slightly higher mass, Fig. 8 suggests the intriguing possibility that the early-G dwarfs of NGC 6819 have depleted more Li than those in the younger Hyades: the only two stars cooler than 6150 K lie below the mean trend of the Hyades, one severely so.

With the inclusion of NGC 6819 \citep{ME15}, the period-mass-age relation was extended beyond the Hyades and NGC 6811 ($\sim$1 Gyr) to an age of 2.3 Gyr, confirming the continued spindown of cooler, lower mass stars with a precision permitting potential age estimation for individual field stars. Moving up the main sequence to higher mass, stellar models that incorporate enhanced angular momentum transport below the convective zone at a level that increases with increasing mass for stars between 0.95 $M_{\sun}$ and 1.15 $M_{\sun}$ can generally reproduce the observed trend of surface rotation and Li abundance with age for stars of different mass \citep{SOP16}. According to the models, for stars just redward of the cool side of the Li dip, the critical factor dominating Li evolution is the convective zone - radiative zone interface, with significant differential rotation early on producing a rapid decline in Li, but giving way to almost solid body rotation as the star ages, producing a flattening of the Li trend for stars older than $\sim$2 Gyr. 

Of particular relevance for the current analysis is the convergence of the period-mass relations for clusters of all ages as the mass of the main sequence star increases, leading to a unimodal, short-period trend for stars with $(B-V)_0$ below $\sim$0.47. In both NGC 6811 and NGC 6819, the rotation period declines precipitously between $(B-V)_0$ = 0.55 and 0.45, the Kraft break \citep{KR67}, but the limiting periods for the two clusters are 1.3 days for NGC 6811 and just under 4.8 days for NGC 6819 \citep{ME15}. If the stars in NGC 6819 had the same rotational distribution as those in NGC 6811 at the same age, they have spun down by almost a factor of 4 over 1.3 Gyr. This decline in rotation speed among the cooler stars in the Li dip may be the continuation of a pattern of gradual spindown exhibited by stars in the Pleiades cluster (100 Myr) and continuing through M35 \citep{GE10}, M34, NGC 2516, to at least the age of the Hyades \citep{TE02, CU17}, i.e. the range in rotational speed among stars with $(B-V)_0$ between 0.47 and 0.55 declines more gradually over time, in contrast with the more rapid change found among stars of lower mass. 
Fig. 8 illustrates that stars in NGC 6819 in this range of $T_{\mathrm{eff}}$ ($6600 - 6300 K$) have clearly depleted more Li, on average, than those in the Hyades.  This additional Li depletion in NGC 6819 may thus be correlated with the additional angular momentum loss these stars have suffered relative to the Hyades. 

The sharp break in the period (rotation speed) -- color (spectral type) relations near $(B-V)_0$ $\sim$ 0.47 is generally attributed to the absence of effective magnetic braking among the hotter stars due to the transition from stars with convective to radiative atmospheres.
Attempts at more realistic 3-D modelling of the convective layer for a star at 1.47 $M_{\sun}$, almost exactly the mass of a star just beyond the high mass edge of the Li-dip boundary, show that a convective overshoot layer is created at the base of the traditional mixing-length model, extending the convective zone deeper and hotter than predicted by more traditional approaches, but still well above the Li-destroying zone \citep{KI16}. Yet rapidly-rotating stars on the hot side of the Li dip have suffered Li depletion. If angular momentum redistribution caused by internal mixing and magnetic braking coupled to a shallow convective atmosphere are the predominant sources of the spindown on the main sequence, stars blueward of the Kraft break \citep{KR67} may simply take longer ($\sim$200 Myr) to develop a more gradual degree of braking from a higher initial rotation rate. 

Another insight about the relationship of angular momentum loss and Li depletion, or lack thereof, comes from short-period, tidally-locked binaries (SPTLBs)]. According to tidal circularization theory \citep{ZA89}, close binaries with periods less than about 8 days would have tidally locked during the early pre-main sequence phase, before the stellar interior was hot enough to destroy any Li. Therefore, SPTLBs could exhibit higher Li abundances than normal single stars at a given phase of evolution \citep{SO90, DE90}. 
Indeed, SPTLBs within the disk and among moderately metal-poor stars \citep{DE97, RY95}, X-ray binaries \citep{MA05}, and V505 Per \citep{BA13}, as well as members of the Hyades \citep{TH93} and M67 \citep{DE94}, 
exhibit higher A(Li) than comparable single stars do. 

A number of caveats should be noted. In SPLTBs other complications could come into play, e.g. meridional circulation in very close binaries, so it is not expected that all SPTLBs would be better preservers of Li than non-binaries. Consistent with theoretical predictions, certain classes of SPTLBs do not show high Li: a) SPTLBs in the Pleiades (100 Myr) are Li-normal, as expected since the Pleiades are too young for rotationally-induced mixing to have become effective in depleting the surface Li; b) binaries with P $>$ 8 days are Li-normal; and c) short-period binaries with mid-F or earlier spectral types are Li-normal, as expected, since such hotter stars would not have been able to tidally lock during the early pre-main-sequence phase \citep{ZA89, RY95}. 
We encourage study of SPTLBs in NGC 6819. 

\subsection{Rotation of Hotter Dwarfs in NGC 6819 and Other Clusters}
Before expounding on what we learn from NGC 6819, we point out varied additional evidence favoring rotationally-induced mixing as the primary cause of the Li dip over other proposed mechanisms.  This evidence includes 
the Li/Be ratio, the Be/B ratio, and the timing of the formation of the Li dip.  Since Li, Be, and B survive to different depths, combined knowledge from two or more of these elements can offer extremely powerful constraints.  For example, if the efficiency of mixing depends on depth in different ways, the resulting surface Li/Be ratio will be affected in different ways (greater shallower mixing will affect Li more than it does Be). \citet{DBS98} found that the depletion of Li and Be in F dwarfs is closely correlated, and the well-defined slope of A(Li) versus
A(Be) strongly favors the predictions of the Yale rotational models \citep{DE97} and rotational models from \citet{Char94}, while strongly arguing against models with diffusion 
\citep{MI86,RM93} and and mass loss \citep{SC90}.
These data also argue against the gravity-wave-induced mixing models of \citet{GA91}, a non-rotational slow mixing mechanism in which Li is depleted more severely relative to Be than in the rotational models. These conclusions have been supported by a number of additional Li/Be studies in field stars and open clusters (e.g., \citet{BDKS, BK02, BAK04}. Boron survives to a greater depth than does Be and can thus provide additional invaluable constraints on the character of the Li-dip-forming mechanism, especially if it is mixing.  As with the Li/Be depletion correlation, B depletions detected in severely Li-depleted and Be-depleted field \citep{BDSL,BO05} and Hyades \citep{BO16} F dwarfs leads to a B/Be depletion correlation that also strongly favors rotational mixing over diffusion, mass loss, and other types of mixing.  Finally, {\it timing} is important: rotational mixing begins closer to the age of the Pleiades whereas diffusion and mass loss become more prominent closer to the age of the Hyades.  Detection of the beginning of the formation of the Li dip in M35 \citep{SD04}, a cluster just slightly older (about 150 Myr) than the Pleiades thus favors rotational mixing; arguably, a few F dwarfs in the Pleiades may
already be depleting their Li.

Fig. 11 presents the rotational velocity distributions among the turnoff stars for four clusters processed and analyzed in the same way, NGC 6819 at 2.25 Gyr, NGC 3680 at 1.75 Gyr \citep{AT09}, NGC 2506 at 1.85 Gyr \citep{AT16, AT18a}, and NGC 7789 at 1.5 Gyr \citep{BR13, RI13, AT19}. The spectroscopic samples for NGC 3680, NGC 2506, and NGC 7789 have all been matched with {\it Gaia} DR2 astrometry to eliminate probable nonmembers, following the same procedure adopted for NGC 6819. Because NGC 2506 is more metal-poor, the mass of the stars occupying the Li dip should be lower than in NGC 3680. Qualitatively, one therefore needs to observe NGC 2506 at a greater age than NGC 3680 to place the stars in the vertical turnoff above the Li dip in the same relative evolutionary phase as a more metal-rich cluster like NGC 3680. In short, despite being older than NGC 3680, from a Li-dip perspective at the turnoff, NGC 2506 is qualitatively younger.

To place the cluster comparison on a common scale while minimizing issues of reddening, metallicity, and distance, stars have been defined in $V$ based upon their magnitude difference relative to $V$ at the high mass (luminous) edge of the Li dip, where $\Delta$$V$ = ($V_{star}$ - $V_{dip}$). Thus, stars with negative values (open black circles) lie at higher mass, outside the Li dip, while positive and increasing $\Delta$$V$ extends across and beyond the cool, low-mass edge of the Li dip (red crosses). For NGC 2506, the spectroscopy did not reach the level of the dip. Using the cluster parameters derived in \citet{AT16} and \citet{AT18a}, coupled with the discussion of the central mass of the Li dip and its relative boundaries in $V$ \citep{AT09}, stars at the center of the evolved NGC 2506 Li dip should have a mass $\sim$1.28 $M_{\sun}$, with the luminous edge positioned $\sim$0.3 mag brighter than the center, placing the predicted cliff in Li abundance for NGC 2506 at $V$ = 15.75, just below the faint limit of the spectroscopic sample. We will adopt this magnitude as $V_{dip}$ for NGC 2506. For NGC 7789, NGC 6819, and NGC 2506 only single-star members are included in the analysis; due to the smaller statistical sample for NGC 3680, all members, binaries or not, were retained. For NGC 7789, one star lies beyond the high-velocity edge of the plot at $V_{ROT}$ = 126 km s$^{-1}$ and $\Delta$$V$ = -0.9. Stars with even higher rotation speeds in NGC 7789 have not been included because the distortion caused by the extreme rotation made reliable estimation of their speeds and their Li abundances impossible. All stars observed in NGC 2506 have been included. Four stars in NGC 3680, KGP 988, 1410A, 1347 and 1506 \citep{KO95}, with measurable rotation speeds between $V_{ROT}$ = 40 and 60 km s$^{-1}$ and $\Delta$$V$ between -0.35 and -1.5 are not plotted. All four stars originally were classed as possible binaries due to the unusual width of the spectral lines \citep{AT09}; if evaluated as single stars with rapid rotation, 988 and 1347 have A(Li) = 3.24 $\pm$ 0.04 and 3.05 $\pm$ 0.06, respectively. If processed as SB2s, the paired stars in 988 and 1410a generate A(Li) = 2.8 $\pm$ 0.1 and 2.75 $\pm$ 0.15, respectively.

Equally important, the lower bound near $V_{ROT}$ $\sim$ 20 km s$^{-1}$ in NGC 7789 is not an artifact of the analysis; red giants in the same cluster with rotational velocities below this limit are readily measured, thus {\it all} stars in our sample rotate faster than 20 km s$^{-1}$.
 
\begin{figure}
\figurenum{11}
\plotone{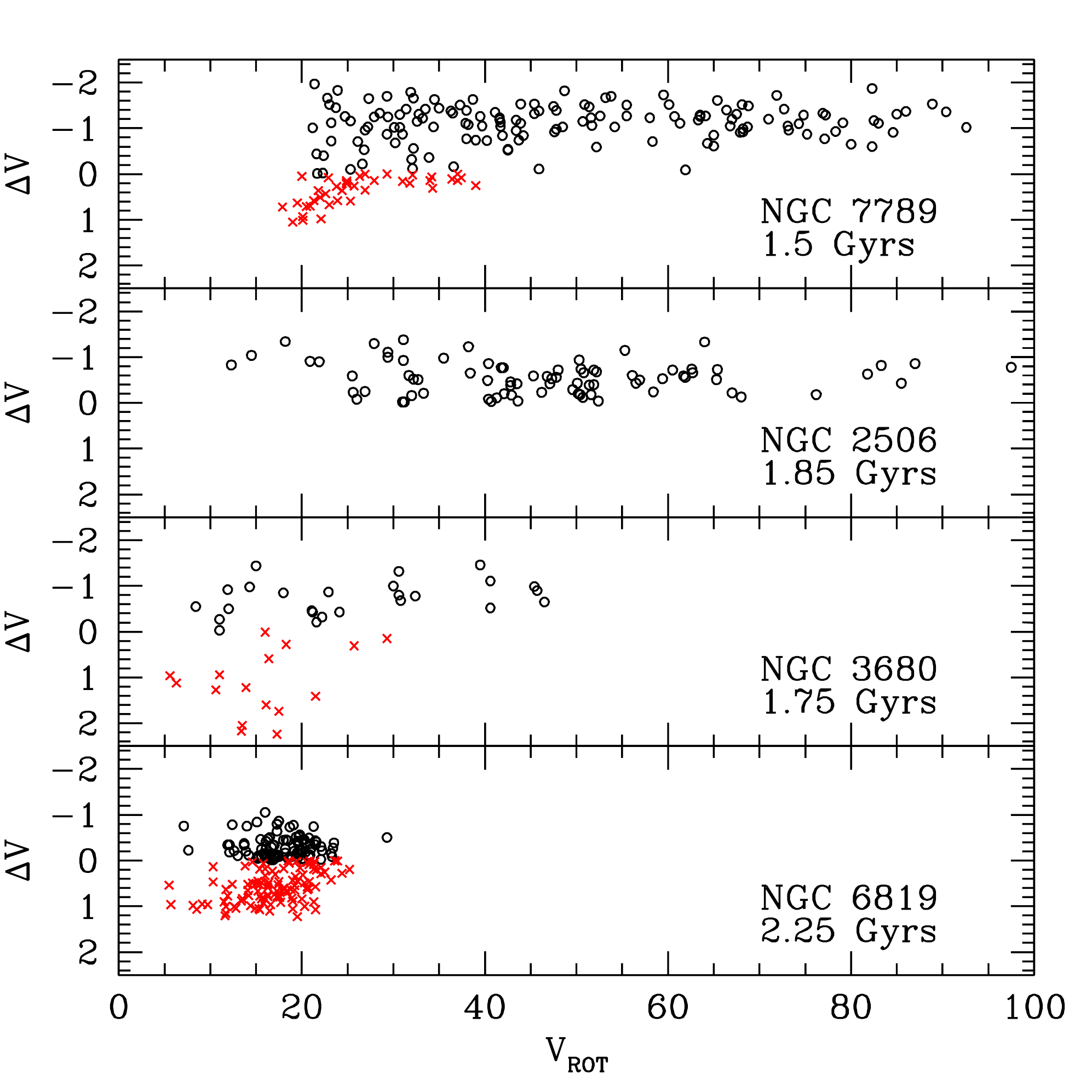}
\caption{Rotational velocity distributions among the stars with measurable Li in the turnoffs of, from bottom to top, NGC 6819, NGC 3680, NGC 2506, and NGC 7789. $\Delta$$V$ = ($V_{star}$ - $V_{dip}$). Stars with negative values (open black symbols) lie at higher mass outside the Li dip, while increasing $\Delta$$V$ extends across and beyond the cool/low mass edge of the dip (red crosses). No stars were observed within the Li dip for NGC 2506. One star in NGC 7789 at $V_{ROT}$ = 126 km s$^{-1}$ and $\Delta$$V$ = -0.9 lies beyond the right edge of the figure.}
\end{figure}

While the turnoff of the younger NGC 7789 extends to higher mass (more negative $\Delta$$V$) than in NGC 2506, the spread in $V_{ROT}$ and the mean $V_{ROT}$ are greater in the younger cluster for stars above the Li dip. The shifts toward lower $V_{ROT}$ in the distribution above and within the Li dip for NGC 3680 and especially for NGC 6819 relative to NGC 7789 are dramatic. By the age of NGC 6819 there is almost no statistically significant difference in the distribution of $V_{ROT}$ between the higher mass stars outside the Li dip and those within, implying that the more rapid rotators among the higher mass stars, as illustrated by the black circles in NGC 2506 and NGC 3680, have spun down to the lower rotation levels of the cooler dwarfs by the age of NGC 6819. For the single-star red giants beyond $(B-V)_0$ = 0.54, the mean $V_{ROT}$ of the NGC 6819 members is cut in half again to values no greater than 8.4 $\pm$ 3.3 km s$^{-1}$.

The relevance of the $V_{ROT}$ pattern for the evolution of Li becomes apparent in Fig. 12, where the distributions of Li among {\it only} the stars brighter than the Li dip (black circles of Fig. 11) are illustrated. To avoid potential issues with the zero-points of the A(Li) scale from one cluster to the next, the A(Li) distribution is based upon $\Delta$A(Li), the difference in A(Li) between a given star and the highest value in a cluster sample. Histograms are based upon the percentage of the stars relative to the total cluster sample above the Li dip. The solid lines represent the fraction of stars with detectable Li in a given $\Delta$A(Li) bin, while the dashed curve is the percentage counting stars with either measured A(Li) or upper limits within the A(Li) range. Thus, the dashed curve should always sit on or above the solid histogram. For NGC 7789, with a  significantly larger sample size than the other three clusters, scatter in A(Li) led to an initial $\Delta$A(Li) bin between 0 and 0.2 with only five stars, creating an artificial offset in the distribution relative to the other three clusters. For comparison purposes, we have reset the upper bound in A(Li) in NGC 7789 to be 0.2 dex lower than the maximum observed and counted the five resulting negative values of $\Delta$A(Li) into the first bin between 0.00 and 0.2.  
    
\begin{figure}
\figurenum{12}
\plotone{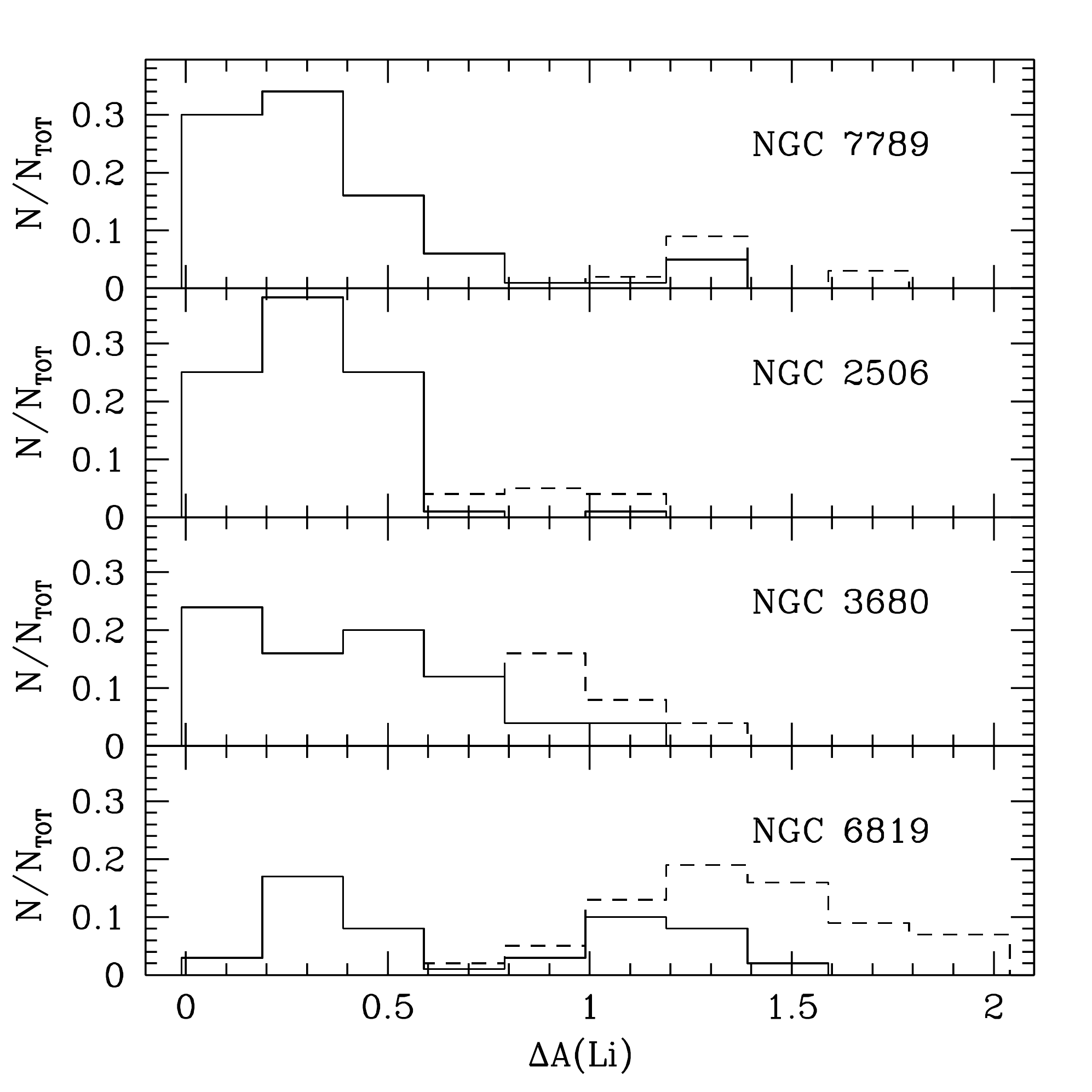}
\caption{A(Li) distribution among the stars hotter/brighter than the Li dip. Solid lines represent the fraction of stars based upon only those with measurable Li while the dashed curve includes stars with upper limits within the A(Li) range. $\Delta$A(Li) equals the difference in A(Li) between a star and the highest measured value within the cluster. The plot sequence is the same as in Fig. 11.}
\end{figure}

Once again, the change from NGC 7789 to NGC 6819 is obvious. The fraction of stars with measurable Li is peaked at low $\Delta$A(Li) values for NGC 7789 and NGC 2506, flattens slightly for NGC 3680, and is almost bimodal for NGC 6819, with a secondary peak near $\Delta$A(Li) = 1.2. Even more apparent, the fraction of stars with upper limits well below the cluster maximum grows significantly. {\it Clearly, in going from NGC 7789 to NGC 6819, the spin-down of these A dwarfs correlates with increasing depletion of lithium.}
\subsection{A Potential Li - $V_{ROT}$ Link}
The stars within the red hook and the phase entering the subgiant branch are well evolved from their main-sequence state, so their current rotation characteristics should not reflect those of stars on the unevolved main sequence. The masses of the typical stars entering the giant branch are 1.9 $M_{\sun}$, 1.65 $M_{\sun}$, 1.75 $M_{\sun}$, and 1.6 $M_{\sun}$ in NGC 7789, NGC 2506, NGC 3680, and NGC 6819, respectively. Of particular relevance for the current discussion is the series of papers \citep{RO02a, RO02b, RO07, ZO12} detailing the rotational characteristics of early type stars from B through late A. \citet{ZO12} found that, unlike the hotter/higher mass stars which showed a clear bimodal distribution of rotational speeds, the late A stars, ranging from 2.5 $M_{\sun}$ to 1.6 $M_{\sun}$, had a unimodal distribution with a well-defined peak tied to the mass. Stars in the 1.6 to 2.0 $M_{\sun}$ range had a broad distribution in $V_{ROT}${\it sin i} which peaked near 125 km s$^{-1}$; correcting for the inclination effect shifted this peak to 145 km s$^{-1}$. The predicted range in the peak over a change of 0.3 $M_{\sun}$ is only 12 km s$^{-1}$, with stars of lower mass spinning slower. While a fraction of the stars in NGC 7789 have $V_{ROT}$ above 100 km s$^{-1}$, making reliable rotational speed and Li estimation impossible, the averages of the observed $V_{ROT}$ distributions in NGC 7789, NGC 2506, and NGC 3680 are less than 1/2 to 1/3 of the  predicted value for the late A stars. A potential factor which could impact the discrepancy between the \citet{ZO12} sample and the clusters is that the rapidly rotating stars in this mass range may have evolved to a temperature which carries them out of the sample range studied by \citet{ZO12}. However, models which include rapid rotation indicate that these stars live longer at higher luminosity, occupying the same region of the CMD as lower mass, slower rotating stars \citep{BR15b}.

For stars in the mass range of interest, \citet{ZO12} conclude that the stars evolve as differential rotators during their entire main-sequence lifetime, with the mean equatorial velocity accelerating during the first third of the star's main-sequence life, then remaining high or slowing down mildly beyond that point. If correct, this analysis implies that the obvious spindown seen through the comparison of NGC 7789 with the late A-star distribution takes place in the limited time between the main-sequence evolution and the subgiant branch, i.e. when the star is near the cool end of the red hook and beyond.

If the turnoff stars of NGC 7789 and the 1.45 Gyr-old cluster NGC 752 \citep{TW15} are evolved equivalents of the late A stars in \citet{ZO12}, then these stars spend the majority of their main-sequence lives with true rotational speeds between 100 km s$^{-1}$ and 250 km s$^{-1}$. Starting with the insightful analyses of \citet{BR15a, BR15b, BR15c}, the impact of a large range of rotational velocities on the interpretation of the ages of star clusters and field stars below 2 Gyr in age has become an extensive and vibrant area of research, in large part driven by the desire to explain the extended main sequences in CMDs of young and intermediate-age clusters of the Magellanic Clouds \citep[e.g.,][]{MA07, MA08, GO09, GI11}. The implication that a spread in initial rotation rates among stars of a given mass within a cluster will lead to a range of mass among giants of the same age fits well with the so-called double clump identified within intermediate-age open clusters \citep{GI99, GI00, GI09, GO14}. The standard interpretations of this feature are tied to the fact that the giants in these clusters lie near the boundary where He-ignition occurs under non-degenerate or degenerate conditions. If the internal structure of the stars of a fixed mass evolving up the giant branch and beyond can be altered by a range in rotation and/or mass loss prior to He-flash or by a range in mass among the stars leaving the main sequence, potentially due again to a range in rotation speeds on the main sequence, the red giant branch will contain He-core-burning stars populating two clumps. Beyond an age of $\sim$2 Gyr, as in NGC 6819, the rapid rotators have minimal impact and the CMD clump returns to the expected appearance for a single-mass population. In addition to rotation \citep{YA13b, NI15, NI16, BA16, LI17, MA18}, the impacts of binaries \citep{YA18}, variable stars \citep{SA18}, stellar age spread \citep{GO11, GO15, GO17}, and convective core overshoot \citep{YA17} have been investigated, with varying degrees of success. Of particular relevance for the current discussion, however, is the study by \citet{WU16}. CMD evolution is studied in the context of an initially rapidly rotating population of B and A stars that slow down over time due predominantly to their evolutionary expansion from the main sequence to the red giant branch. For the populous Magellanic Cloud clusters, this eliminates the contradiction between a large color spread among the stars at the turnoff feeding into a subgiant branch with a narrow range in luminosity \citep{LI14}. With an age of 1.5 Gyr, the highest mass star above the Li dip in NGC 7789 has already spun down by about a factor of two. 

From Fig. 12, the fraction of turnoff stars brighter than the Li dip in NGC 6819 with detectable Li is comparable to those with only upper limits, in distinct contrast with the other 3 clusters. Closer 
examination of the NGC 6819 pattern for these stars in Fig. 5 reveals that, {\it in comparison with the samples in NGC 7789, NGC 2506 and NGC 3680, the NGC 6819 turnoff stars are 
developing a second Li dip at higher mass than the F-dwarf Li dip, with the greatest concentration of stars with detectable Li located just above the  high mass boundary of the Li dip.} Since the brighter stars above the Li dip in NGC 6819 populate the red hook, it is tempting to assume that the growth of the convective zone as the stars expand into the red hook and beyond is the primary source for this Li depletion. This argument fails, however, when applied to NGC 7789, NGC 2506, and NGC 3680. 

As noted earlier, the striking feature about the blue edge of the Li dip is the rapid change in the Li-depletion process as one changes the mass ($T_{\mathrm{eff}}$) of the main-sequence stars by a very small amount. If $T_{\mathrm{eff}}$ were the sole determinant of the effectiveness of the process driving Li-depletion, as soon as stars above the Li dip evolved redward off the main sequence toward a vertical turnoff, they should trigger the same process which depletes Li among stars within the Li dip. As an example, using the VR isochrones and Fig. 10, stars defining the high mass edge of the Li dip have a mass of 1.42 $M_{\sun}$. Tracking back to the unevolved main sequence as defined by the CMD location of the boundary at the age of the Hyades, these stars had a temperature of approximately 6900 K. Evolution off the main sequence moves this boundary to 6870 K at 0.9 Gyr, 6830 K at 1.2 Gyr, 6760 K at 1.5 Gyr (the age of NGC 752 and NGC 7789), 6680 K at 1.75 Gyr (NGC 3680), and 6480 K by 2.25 Gyr (NGC 6819). Between the age of the Hyades when the Li dip is fully developed and the age of NGC 6819, the stars defining the hot boundary evolve in $T_{\mathrm{eff}}$ across a range which encompasses the entire Li dip. By the age of NGC 7789, stars at the hot boundary have spent at least 0.3 Gyr within the Li dip and all stars more massive than this boundary have been there longer. Despite this, the large majority of stars above the turnoff in NGC 7789, NGC 2506, and NGC 3680 still retain A(Li) near their supposedly original values. Even at the age of NGC 6819, a significant fraction of the stars at the higher-mass boundary still have detectable and high values of A(Li).

As discussed in detail for NGC 2506 \citep{AT18a}, a plausible answer to this delayed reaction comes from the mechanisms commonly used to explain the Li dip itself. The depth of the convection zone among stars hotter than the Li dip ranges from nonexistent to inadequate for driving Li depletion. As demonstrated by the models of \citet{CH10}, inclusion of significant rotation and rotationally-induced mixing can produce an immediate and continuous decline in A(Li) as stars evolve across the subgiant branch, especially in light of the deepening convective zone at cooler $T_{\mathrm{eff}}$. The pattern revealed in the turnoff regions of the clusters under discussion implies that Li depletion only becomes significant for stars entering the main sequence red hook if the time spent in this phase, the growth of the convective zone, and the degree of rotational spindown combine to induce a serious depletion in the atmospheric Li level. A high degree of depletion inevitably occurs for the majority of stars evolving along the giant branch, but the initiation point for significant depletion shifts to earlier phases of post-main-sequence evolution and lower mass as a cluster ages. 
\subsection{Evolved Stars: Li in Giants and Subgiants}

The striking change in the A(Li) distributions for the turnoff stars in going from NGC 7789 to NGC 6819 is reflected in the giants, though the pattern is somewhat modified by the relative change in the evolutionary phases occupied by post-main-sequence stars with increasing age. As discussed earlier, the transition from NGC 7789 to NGC 6819 is important in part because it covers the mass and age range where stars leaving the main sequence change from nondegenerate to partially degenerate cores, and He-ignition at the tip of the giant branch switches from a quiescent phase to He-flash under degenerate conditions. In fact, a mixture of red giants undergoing both forms of He-ignition within the same cluster has long been a potential explanation for the peculiar red giant clumps in both NGC 7789 and NGC 752 \citep{GI00}, as discussed earlier. 

The state of the star in the hydrogen-exhaustion phase determines the rate of evolution across the subgiant branch and up the first-ascent red giant branch. The state of the star at He-ignition should seriously impact the degree of mixing, raising the prospect of serious Li-depletion among stars undergoing He-flash, i.e., most if not all red clump stars should exhibit greatly reduced Li abundances. 
Testing this idea is complicated.  For example, nearly all stars on the first ascent RGB in NGC 6819 already have unmeasurably low A(Li), making it impossible to know whether there are further reductions in A(Li) after the He flash on the way to the red clump. These low A(Li) values in NGC 6819 are largely due to Li depletion during the MS, so the distribution of A(Li) in all phases from the MS through the helium flash must be taken into account.
The measured Li distribution among the subgiants and giants will depend upon the relative population of subgiants, first-ascent red giants and red clump stars, as well as the masses of the stars feeding the giant branch. To understand the evolutionary impact, a simple comparison of NGC 7789 and NGC 2506 is in order. As seen in Fig. 12, 87\% and 88\% of the stars at the turnoff above the Li dip in NGC 7789 and NGC 2506, respectively, have measurable Li within 1 dex of the cluster maximum. Since A(Li) is easier to measure among stars cooler than this color range, any decline in the fraction of subgiant and giant stars with comparable A(Li) must be a reflection of a real reduction in the elemental abundance. However, as shown in the CMD distribution for clusters with distinctly different ages, the subgiant branch and the giant branch below the clump in the younger NGC 7789 are sparsely populated \citep{GI98, TW12, BR13} while the comparable regions in the older NGC 2506 are easily delineated \citep{AT16}, a primary factor in its selection for Li analysis to test the degree of mixing occurring across the subgiant branch \citep{AT18a}. Thus, the giants in NGC 7789 are almost totally dominated by stars at or above the level of the red giant clump while in NGC 2506, the majority of stars are first-ascent red giants below the level of the clump. As expected for more evolved stars, both clusters show a reduction in the percentage of red giants with measurable Li, dropping to one-third for NGC 7789 and one-half for NGC 2506, with the greater drop for NGC 7789 tied to the more advanced evolutionary state of the giants in that sample.

By contrast, among the single stars within the giant branch of NGC 6819 (Fig. 10), only 10\% have measurable A(Li), despite a first-ascent giant branch as well populated as that within the younger NGC 2506. The key is that, in addition to any Li depletion which might occur across the subgiant branch or at the first dredge-up, the stars leaving the main sequence are already reduced to a supply of only 35\% with A(Li) measurable within 1 dex of the cluster maximum.

The majority of giants within NGC 3680 do have measurable A(Li), but the significance is minimal given a sample of only nine stars, four of which are binaries \citep{AT09}.

\section{Summary and Conclusions}
As emphasized in Paper II, for any attempt to delineate the evolution of Li with stellar age and mass, NGC 6819 displays a rare blend of critical characteristics. It is young enough for the turnoff stars to cover the full range of mass defining the Li dip, with the higher mass stars above the Li dip still retaining limiting values that should be representative of the initial cluster abundance, but old enough that the stars evolving through the subgiant branch and ascending the giant branch for the first time have partially degenerate cores, leading ultimately to He-flash at the tip of the giant branch. Despite an age significantly larger than the typical open cluster evaporation timescale, the cluster is rich in members bright enough to allow high dispersion spectroscopic analysis below the level of the Li dip. The precision radial-velocity, proper-motion, and photometric surveys of the cluster field have greatly enhanced identification of probable cluster members while allowing for detection of and correction for variable reddening across the face of the cluster, a necessity for reliable temperature determination as a basis for spectroscopic and photometric analysis. Finally, the inclusion of the cluster within the {\it Kepler} field allows asteroseismic tests  of the internal structure and evolutionary phase for the evolved cluster members, confirming in the case of star 7017 the anomalous nature of this star relative to the other cluster members.

Building upon the precision photometry and spectroscopy of Papers I and II, as well as the current abundance re-evaluation from ANNA, it is found that for a mean cluster reddening in the range of $E(B-V)$ = 0.14 to 0.16, having corrected for variable reddening across the cluster face and adopting a slightly subsolar mean metallicity of $[Fe/H] = -0.04$, the best estimate for the cluster age is 2.40 to 2.25 Gyr, with an apparent distance modulus between 12.29 and 12.40, in excellent agreement with the zero-point-adjusted parallax measures from {\it Gaia} DR2. Ages closer to 2.6 Gyr can be obtained by adopting artificially lower reddening or by selecting different sets of isochrones, but the latter alteration is a reflection of the range among theoretical models rather than a problem with the observational data for the cluster. It should be noted that adoption of the older spectroscopic abundance of $[Fe/H] = +0.09$ for the cluster \citep{BR01} would reduce the age below the currently derived value of 2.25 to 2.4 Gyr.

Turning to Li, the limiting value of A(Li) among the single stars with the highest mass at the turnoff is A(Li) = 3.2 $\pm$ 0.1, consistent within the errors with that found for the primordial solar system. The majority of the stars with A(Li) above 3.2 are within binary systems, potentially indicating that they have retained their primordial Li value while the remaining single stars, having evolved well off the ZAMS and now approaching the hydrogen-exhaustion-phase (HEP) and beyond, may have undergone Li depletion to varying degrees. The depletion range in A(Li) among the single turnoff stars above the high-mass boundary of the Li dip is as large as that found within the dip.

For stars on the low-mass side of the Li dip, the mean A(Li) = 2.83, with no trend with magnitude over the range from $V_0$ = 15.5 to 15.9. Equally important, the dispersion in A(Li) among these stars is 0.16 dex, three times larger than expected from spectroscopic errors alone. So, while there are stars fainter than the Li dip with A(Li) approaching the limiting value for stars on the hot side of the Li dip, the significant lack of differential evolution predicted among the cooler stars implies that these stars merely represent the high-Li end of an intrinsic scatter centered near A(Li) = 2.85. In short, the cooler stars have depleted their initial Li from A(Li) $\sim$ 3.3 to the current mean of 2.83, with a scatter tied to some intrinsic property of the sample.

The Li dip within the vertical turnoff of the cluster, when translated back to the age of the Hyades/Praesepe and adjusted for a modest shift in the $T_{\mathrm{eff}}$ scale, shows a profile with $T_{\mathrm{eff}}$ that bears a striking resemblance to that defined by the younger clusters. On the hot or high-mass side of the profile, the transition from measurable and high Li to depleted or unmeasurable Li takes place over a ZAMS $T_{\mathrm{eff}}$ range of less than 80 K. On the hot side above the Li dip, the limiting value of A(Li) for NGC 6819 lies below the limiting estimate for the much younger Hyades for the reasons noted previously. Taking this into account, there is little evidence for a widening in the mass or $T_{\mathrm{eff}}$ range of the Li dip between 0.6 Gyr and 2.3 Gyr. The stars on the cool slope of the Li dip may show some depletion in NGC 6819 relative to the Hyades, but the cool boundary of the Li dip is an excellent match to that within the Hyades. Within the center of the Li dip, solid conclusions are more difficult to achieve because the majority of stars within this mass range have only upper limits to A(Li).

Of the 65 stars with $(B-V)_0$ greater than 0.54 on the subgiant branch and beyond, only 8, including 2 binaries, have detectable Li and of these, only 3, including 7017, overlap with the range of detectable Li among the stars at the turnoff above the Li dip. The implied pattern of a significant depletion of Li among stars just leaving the main sequence but well before the the phase of the first dredge-up is consistent with the need for an additional mixing mechanism among stars with masses in the range leading to partially degenerate cores after the HEP, as illustrated by IC 4651, NGC 752, NGC 3680  \citep{PA04, AT09} and NGC 2506 \citep{AT18a}. The greater cluster age and lower turnoff mass also correlates with the absence of an apparent double clump associated with clusters like NGC 752 and NGC 7789 and is more representative of the pattern in the older cluster M67, where stars leaving the main sequence emerge from within the Li dip \citep{PA12}.

As noted, while the boundary of the Li dip on the high-mass edge is sharply defined, stars above this edge display a significant range in Li, from an approximately primordial solar value to upper limits competitive with those found within the Li dip. Equally important, the distribution of A(Li) among these stars is heavily skewed toward values an order of magnitude lower than the cluster limiting/primordial value, in striking contrast with younger clusters like NGC 7789 and NGC 2506 where the majority of stars have measurable A(Li) within 1 dex of the cluster limit. In short, as the stars higher in mass than the Li dip evolve through the HEP, they are reproducing the pattern of depletion found within the Li dip itself, but at a much later phase of evolution. It should be reemphasized, however, that the mechanism defining the Li dip cannot be purely $T_{\mathrm{eff}}$-dependent for the stars of higher mass; the stars more luminous than the Li dip boundary enter the $T_{\mathrm{eff}}$ range defining the Li dip on the unevolved main sequence long before significant depletion becomes detectable, in stark contrast with the traditional Li dip which exhibits depletions in clusters by 100 Myr and is fully formed by the age of the Hyades/Praesepe \citep{CU17}.

Adjusting for the fact that the mass ranges among the turnoff stars within NGC 7789 and NGC 2506 extend to higher values than in NGC 6819, when comparing stars of comparable relative mass above the Li dip, the mean and dispersion of $V_{ROT}$ among the stars in NGC 7789 and NGC 2506 are larger by a factor of 2-3 than among the stars in NGC 6819. Moreover, the mean and dispersion of $V_{ROT}$ among the stars above and below the high-mass boundary of the Li dip at the turnoff of NGC 6819 are essentially identical. When coupled with the minimization of Li-depletion among SPTLBs, it is concluded that the second critical factor controlling mixing and Li-depletion appears to be the rate of stellar spindown, rather than the absolute size of $V_{ROT}$.

Within the greater context of Li-depletion along the entire main sequence, the observations indicate that the Li dip potentially arises as a consequence of the convolution of two mass-temperature-dependent functions. Moving along the main sequence from low to high mass, the average depth of the surface convection zone grows shallower, leading to slower rates of Li-depletion until stars like the sun and higher mass should exhibit little if any Li-depletion in their atmospheres. The second, competing factor is stellar spindown which potentially induces mixing at the base of the convection zone and/or within the atmosphere itself. For stars in the 0.6 to 1.0 $M_{\sun}$ range, the initial bimodal distribution of $V_{ROT}$ converges by $\sim$0.4 Gyr to a unimodal profile dominated by slow rotators. Observations of stars with masses just below solar show a clear trend of decreasing magnetic field strength with age up to the age of the Hyades \citep{FO18}, while analysis of single-star solar analogs shows a clear correlation between declining $V_{ROT}$ and declining A(Li) \citep{BE17}. However, between $(B-V)_0$ = 0.55 and 0.4, the trend of $V_{ROT}$ with increasing $T_{\mathrm{eff}}$ undergoes a sharp rise, and a rapid decline in $V_{ROT}$ by 0.4 Gyr is no longer applicable. For stars on the cool side of the Li dip and just beyond, the initial spindown does occur, but it takes longer and starts later, generating the red edge of the Li dip in lower mass stars between 0.2 and 0.6 Gyr. Thus, even lower mass, solar analogs deplete their surface Li by a factor of 10 over the first Gyr. The spindown-driven mixing and the convection zone continue to deplete Li at a slow but detectable rate over the lifetime of the star \citep{TU15}, leading to the Li-plateau among cooler dwarfs, without reaching deep enough to destroy Be. Clearly, this empirical explanation for the observed pattern remains qualitative at best, in part because current theoretical stellar models which attempt to reproduce the observed spindown among solar-type stars as a function of age using a variety of angular momentum transport mechanisms can do so, but not without failing to simultaneously reproduce the Li abundance and the internal rotational structure of the sun \citep{AM16}. Somewhat surprisingly, the best models for reproducing the evolution of the rotation rate with age often deplete too much Li over time.

Among the hotter stars in the Li dip, the significantly higher initial $V_{ROT}$ distribution ultimately compensates for an even shallower convection zone, driving the mixing zone deep enough to cause Li-depletion and Be-depletion when brought into contact with the thin convective layer of the atmosphere. As the rotation rate slowly declines, the level of rotational mixing subsides and the convection zone no longer maintains contact with the depleted layers below the atmosphere, leaving A(Li) at a fixed level until evolution beyond the main sequence and subsequent spindown during the HEP drive the mixing layer down to where additional Li-depleted gas can be accessed.

\acknowledgments
Extensive use was made of the WEBDA\footnote{http:// webda.physics.muni.cz} database, maintained at the University of Brno by E. Paunzen, C. Stutz and J. Janik.
NSF support for this project was provided to BJAT, DLB and BAT through NSF grant AST-1211621, and to CPD through NSF grant AST-1211699. The authors gratefully acknowledge the thoughtful comments of the referee which led to greater clarity in the extensive discussion of the text.

\facility{WIYN:3.5m}
\software{IRAF \citet{TODY}, MOOG \citet{SN73}, ANNA \citet{LB17,LB18a}}

\clearpage
\end{document}